\PassOptionsToPackage{x11names}{xcolor}
\documentclass[12pt,a4paper,twoside]{article}
\usepackage{array}
\usepackage{amsmath}
\usepackage{tabularx}
\usepackage{txfonts}
\usepackage{enumerate}
\usepackage{float}
\usepackage{xspace}
\usepackage[x11names]{xcolor}
\usepackage{multirow}

\usepackage{booktabs}

\usepackage{subcaption}

\usepackage{pdflscape}

\usepackage[T1]{fontenc}
\usepackage[cp1250]{inputenc}

\usepackage[margin=2cm]{geometry}

\usepackage[colorlinks=true,citecolor=red,linkcolor=Green4]{hyperref}

\usepackage{changes}

\usepackage{listings}
\colorlet{cpop}{Chartreuse3}
\colorlet{cmonika}{DarkOrchid1}
\colorlet{ckonrad}{cyan}
\colorlet{cagata}{DeepPink1}

\definechangesauthor[name={Konrad}, color=ckonrad]{Konrad}
\definechangesauthor[name={Monika}, color=cmonika]{Monika}
\definechangesauthor[name={Agata}, color=cagata]{Agata}
\definechangesauthor[name={POP}, color=cpop]{POP}

\usepackage{pgfplots}
\pgfplotsset{compat=newest}

\usepackage{tikz}
\usetikzlibrary{spy,patterns}
\usepgfplotslibrary{statistics}

\pgfkeys{
	/pgf/number format/.cd,
	sci generic={mantissa sep=\times,exponent={10^{#1}}}}

\definecolor{blues1}{RGB}{236,231,242}
\definecolor{blues2}{RGB}{208,209,230}
\definecolor{blues3}{RGB}{166,189,219}
\definecolor{blues4}{RGB}{116,169,207}
\definecolor{blues5}{RGB}{54,144,192}
\definecolor{blues6}{RGB}{5,112,176}
\definecolor{blues7}{RGB}{3,78,123}
\numberwithin{equation}{section}

\begin{document}
\pagestyle{myheadings}
\markright{Lonc, A., Piotrowska,~M.J.\& Sakowski, K.: Analysis of the hospital records}

% ------------------------------------------------------------------------- %
\noindent\begin{tabular}{|p{\textwidth}}
	\Large\bf Analysis of the hospital records from AOK Plus
 \\\vspace{0.01cm}
    \it Lonc, A.$^{\dagger,1}$, Piotrowska, M.J.$^{\dagger,2}$ \& Sakowski, K$^{\dagger,*,\diamondsuit,3}$\\\vspace{0.02cm}
\it\small $^\dagger$Institute of Applied Mathematics and Mechanics,\\
\it\small University of Warsaw, Banacha 2, 02-097 Warsaw, Poland\\\vspace{0.01cm}
\it\small $^*$Institute of High Pressure Physics,\\
\it\small Polish Academy of Sciences, 01-142 Warsaw, Sokolowska 29/37, Poland\\\vspace{0.01cm}
\it\small $^\diamondsuit$ Research Institute for Applied Mechanics,\\
\it\small Kyushu University, Kasuga, Fukuoka 816-8580, Japan\\\vspace{0.01cm}
\small  $^1$\texttt{alonc96@gmail.com}, $^2$\texttt{monika@mimuw.edu.pl}, $^3$\texttt{konrad@mimuw.edu.pl}\\
    \multicolumn{1}{|r}{\large\color{orange} }\\%Data Report} \\
	\\
	\hline
\end{tabular}
% ------------------------------------------------------------------------- %
\thispagestyle{empty}

%\tableofcontents
%\addtocontents{toc}{\protect\setcounter{tocdepth}{3}}
%\noindent\begin{tabular}{p{\textwidth}}
%	\\
%	\hline
%\end{tabular}
%\vspace{2em}\\

\begin{abstract}
	We present analysis of anonymised admission/discharge data from insurance provider for Saxony and Thuringia (Germany) for years 2010--2016. Study of such data are necessary to derive a structure of healthcare system transfer network, as no patients' transfer data are currently available. Hospital network can be directly used as a basis for modelling of multidrug-resistant pathogen spread allowing to study the effectiveness of disease-control strategies. In this paper, the properties of the dataset under consideration are presented and discussed. 
\end{abstract}

\textbf{Keywords:}  
%Network modelling, multidrug-resistant Enterobacteriaceae, transmission dynamics, infection control,
%healthcare network
healthcare data analysis, overlapping data, healthcare network, multidrug-resistant bacteriaceae
% healthcare associated infections

\section{Introduction}

Multidrug-resistant (MDR) bacteriae recently became more serious health threat \cite{NewEnglJMed-2010-362-1804,JAntimicrobChemother-2008-62-1422}.  According to the
European Centre for Disease Prevention and Control and the European Medicines Agency, MDR bacteriae are responsible for about 25\,000 deaths of patients per year,~\cite{UErep2009}. In addition, it is estimated that these infections result in extra healthcare costs  of at least EUR 1.5 billion each year.
These pathogens can spread within healthcare system population, e.g. by contact between undiagnosed infectious patients and susceptible patients. Pathogen control strategies exist (see e.g. \cite{InfectControlHospEpidemiol-2011-32-1064,JAntimicrobChemother-2008-62-1422}), but they mainly focus on the individual healthcare facility level, while cooperative approach may be more beneficial \cite{Lee2012}. 
Unique properties of MDR bacteriae make them immune to typical prevention strategies, mainly due to their antibiotic-resistant nature. In order to understand how to inhibit spreading of such bacteria, new models describing their transmissions are introduced. Such attempts to model the spread of some particular bacteria within the hospital network has already been attempted, see e.g.~\cite{Donker2012,AmerJPublHealth-2011-101-707,Donker2017}.

Clearly, to model such phenomenon, one needs to have access to a database, providing data of hospital transfers. Unfortunately, such databases are not available even in Europe. Hospitals, or healthcare facilities in general, collect information about patient admissions, discharges, age, sex, diagnosis, procedures in they own patient registers. However hospitalisation history of patients is not collected by healthcare facilities. Nevertheless, an excellent source of data could be the healthcare providers collecting patient hospitalisation records for years. They store basic information about the patients and their healthcare history. Unfortunately, transfer details are not among them.
So, one of the primal difficulties for the researchers would be to derive the transfer data from the hospitalisation history. Despite such task seems to be not too hard at first glance, in fact it can be quite challenging. In addition, more information about a healthcare system structure is necessary for accurate simulation results, for example to divide patients into risk groups. To provide as much important information for the future work concerning the dynamics of pathogen spread within the inter-hospital network, we perform analysis of the dataset provided by AOK Plus -- one of the largest health insurance company in Germany.

This paper is organized as follows. In Section~\ref{sec:description} and~\ref{sec:tools} we briefly describe the database provided by AOK Plus and tools used to analyse data. Next, in Section~\ref{sec:dat:anal}, we present results of our analysis. Finally, in Section~\ref{sec:summary}, we summarise and discuss obtained results.

\section{Description of dataset}\label{sec:description}
We consider anonymized patients dataset provided by AOK Plus – a healthcare provider in Saxony and Thuringia. Dataset consist of 4\,826\,823 hospitalisation records of 1\,623\,567 patients covering period of 7 years (2010 -- 2016). In particular, the database stores the following information: patient anonymized ID, anonymized healthcare facility ID, federal state of healthcare facility, day of the admission, day of discharge, diagnosis (international ICD-10-GM code), patient's sex and year of birth. 

Within provided dataset we have found 2\,991\,597 hospital/healthcare facility stay records for the facilities located in Saxony with the numeric diagnosis code, 1\,566\,451 for Thuringia, 268\,182 for other German federal states and 593 records without any location given (for more details see Table~\ref{tab:des:location}). There are 1\,925 unique hospital facilities among the whole database and 134 of them are situated in Saxony and Thuringia.

\begin{table}
	\centering
	\caption{Number of admissions for given locations \label{tab:des:location}}
	\begin{tabular}{|r|r|}
		\hline
		{\bf Location}&{\bf Number of records}\\ \hline
		Saxony  & 2 991 597 \\ \hline
		Thuringia  & 1 566 451 \\ \hline
		Bavaria  &   76 099 \\ \hline
		Saxony-Anhalt &   41 881 \\ \hline
		Brandenburg  &   29 219 \\ \hline
		Lower Saxony  &   26 759 \\ \hline
		Hesse  &   25 008 \\ \hline
		Berlin  &   18 592 \\ \hline
		North Rhine-Westphalia  &   15 364 \\ \hline
		Baden-Württemberg  &   13 053 \\ \hline
		Mecklenburg-Vorpommern  &   8 383 \\ \hline
		Rhineland-Palatinate &    5 045 \\ \hline
		Schleswig-Holstein  &    4 045 \\ \hline
		Hamburg  &    3 017 \\ \hline
		Bremen  &    906 \\ \hline
		Saarland  &     811 \\ \hline
		no location&    593 \\ \hline
	\end{tabular}
\end{table}

\section{Data analysis tools}\label{sec:tools}

From the admission and discharge data, we determine the duration of patent stays for each hospitalisation record. To determine the history of each patient, we group the records with the same patient identification number. However, similarly as in our previous study of the AOK Lower Saxony dataset~\cite{Piotrowska2019}, we faced the problem of so-called \emph{overlaps}, existing in the provided dataset (for the definition see Subsection~\ref{sec:ovelaps}). To investigate the provided dataset (admissions, discharges, duration of stays, sizes of the facilities etc.) and to detect the overlaps, analyse their structure and statistics, we use previously developed Python code, freely available (with documentation) on the web page \url{www.mimuw.edu.pl/~monika/emergenet}. 
 
\section{Data analysis}\label{sec:dat:anal}

\subsection{Population structure}
Within all the records we found data for 741\,346 men with 1 up to 198 hospitalisations and 882\,221 women with 1 up to 148 hospitalisations. The average number of admissions per person is 3.09 for men and slightly lower 2.88 for women, however the median of number of admissions per person is 2 independently of sex. The average length of patient hospitalisation is 9.8 days, and the average period spend outside the facilities, between two hospitalisations is 284.9 days.

\begin{figure}
	\centering
	\includegraphics[width=0.9\linewidth]{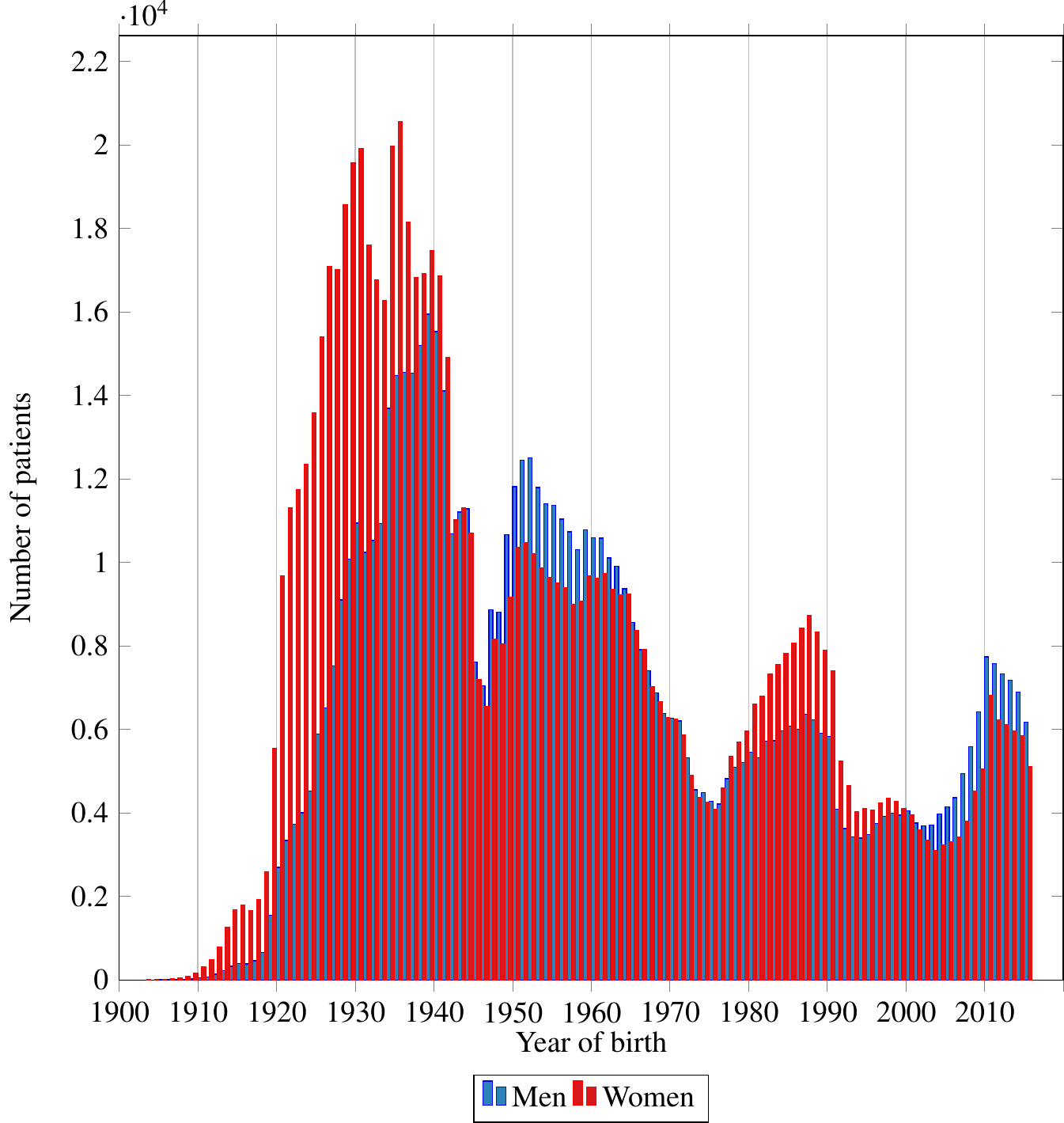}
%	\begin{subfigure}[t]{\textwidth}
%		\begin{tikzpicture}
%		\begin{axis}[
%		ybar,
%		width = \textwidth, height = 0.3\textwidth,
%		bar width = 0.4,
%		/pgf/number format/1000 sep={},
%		legend style={at={(0.5,-0.30)},anchor=north,legend columns=-1},
%		ymin=0,
%		xmin=1900,
%		xmax=2016,
%		ylabel={Number of patients},
%		xlabel={Year of birth},
%		]
%		\addplot table [col sep=semicolon] {wyniki_text/byear_stat/byear_stat.csv};
%		\end{axis}
%		\end{tikzpicture}
%		\caption{}
%	\end{subfigure}%
%%	~
%%	\begin{subfigure}[t]{0.5\textwidth}
%%		\begin{tikzpicture}
%%		\begin{axis}[
%%		ybar,
%%		width = \textwidth, height = \textwidth,
%%		bar width = 0.2,
%%		/pgf/number format/1000 sep={},
%%		legend style={at={(0.5,-0.30)},anchor=north,legend columns=-1},
%%		ymin=0,
%%		xmin=1900,
%%		xmax=2016,
%%		ylabel={Number of patients},
%%		xlabel={Year of birth},
%%		]
%%		\addplot+ table [col sep=semicolon] {wyniki_text/byear_stat/st_byear_stat.csv};
%%		\end{axis}
%%		\end{tikzpicture}
%%		\caption{}
%%	\end{subfigure}
	\caption{Structure of the patients' population considering data from all healthcare facilities within years 2010-2016.\label{fig:popul}}
\end{figure}

In Figure~\ref{fig:popul} we present the structure of patients population in the database. Clearly, we do not consider patients age, as the database covers seven years. Furthermore, we do not know exactly which patients died during that period -- such information is only available in the database if a patient died while being insured by AOK Plus. Thus, for our statistical purposes we consider the birth year to investigate the age structure of patients.

Among 4\,826\,823 hospitalisation records there are 886\,650 (18.4\%) cases of diseases of circulatory system. Other significant groups are diseases of the digestive system (488\,095, 10.1\%), injury, poisoning and certain other consequences of external causes (487\,487, 10.1\%), neoplasms (487\,008, 10.1\%) and mental and behavioural disorders (352\,759, 7.3\%) (cf. Table~\ref{tab:popul:icd}).

After limiting the records to the facilities located in Saxony and Thuringia, we end up with data for 706\,827 men  with 1 up to 155 hospitalisations and 845\,666 women with 1 up to 147 hospitalisations. The average number of admissions per person is 3.24 for men and 3.0 for women, the median of admissions per person is 2 independently of sex. The average length of hospitalisation is 9.9 days, and the average period spend outside the facilities, between two hospitalisations is 286.3 days. The structure of patients is very similar to the structure for the whole set -- difference between number of patients born in one year in the whole datasets and in Saxony and Thuringia constitutes at most 10.45\% of the former number. As presented in Table~\ref{tab:popul:icd}, the illnesses types are also similarly distributed in both datasets.

\begin{table}
	\centering
	\caption{Number of admissions in the whole dataset and set limited to Saxony and Thuringia for given chapter number of the ICD-10-GM illness code
		(\url{https://www.dimdi.de/static/de/klassifikationen/icd/icd-10-gm/kode-suche/htmlgm2019/}).}\label{tab:popul:icd}
	\begin{tabular}{|r|c|c|c|c|c|c|c|c|}
		\hline
		{\bf Chapter}&1&2&3&4&5&6&7&8\\ 
		\hline
		{\bf Whole dataset}&180050&487008&46233&174666&352759&207178&89796&32294 \\ 
		\hline
		{\bf Saxony, Thuringia} &170705&464666&44631&165296&327221&193478&82988&30487\\
		\hline
		\hline
		{\bf Chapter}&9&10&11&12&13&14&15&\\ 
		\hline
		{\bf Whole dataset}&886650&318726&488095&75047&333196&253954&81395& \\ 
		\hline
		\textbf{Saxony, Thuringia} &843512&302471&466739&70299&308986&241627&74229&\\
		\hline
		\hline
		{\bf Chapter}&16&17&18&19&20&21&22&\\ 
		\hline
		{\bf Whole dataset}&34797&18211&244667&487487&2&34611&1& \\ 
		\hline
		\textbf{Saxony, Thuringia} &31721&15273&231273&460118&2&32325&1&\\
		\hline
	\end{tabular}
\end{table}
\subsection{Admissions}
\begin{figure}
	\centering
	\begin{subfigure}[t]{\textwidth}
			\includegraphics[width=0.9\linewidth]{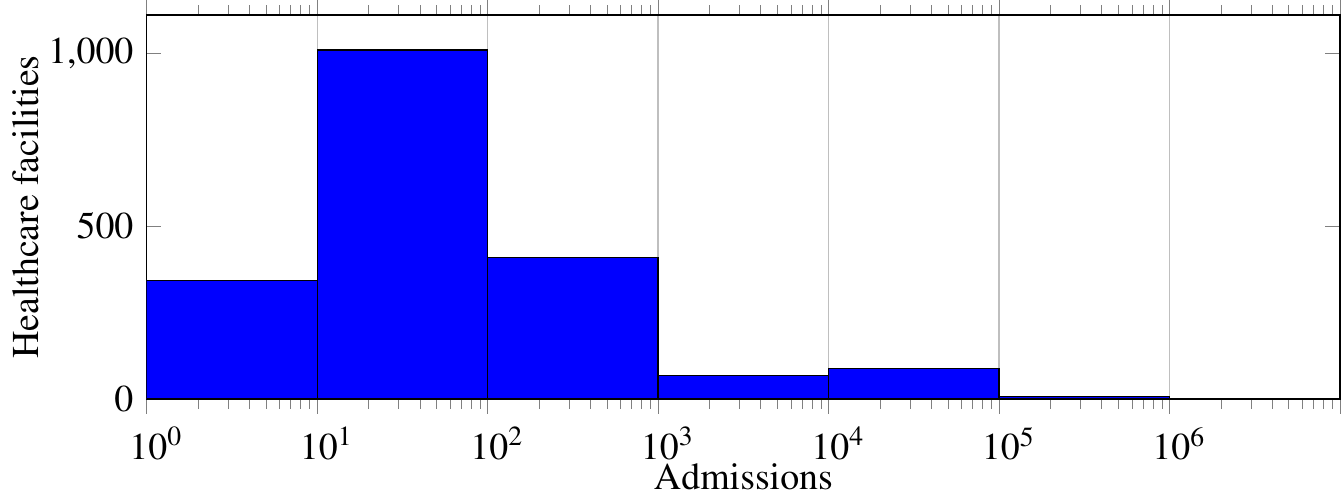}
%		\begin{tikzpicture}
%		\begin{axis}[
%		ybar interval,
%		width = \textwidth, height = 0.4\textwidth,
%		legend style={at={(0.5,-0.30)},anchor=north,legend columns=-1},
%		ymin=0,
%		xmin=1,
%		xmax=10e6,
%		xmode=log,
%		xlabel={Admissions},
%		ylabel={Healthcare facilities},
%		point meta=rawx,
%		xticklabel style={xshift=-2em},
%		]
%		\addplot[fill=blue] table [col sep=semicolon] {wyniki_text/hospitals_entriesno/hospitals_entriesno2.csv};
%		\end{axis}
%		\end{tikzpicture}
		\caption{\label{fig:hosp:entries:all:a}}
	\end{subfigure}
	~
	\begin{subfigure}[t]{\textwidth}
		\includegraphics[width=0.9\linewidth]{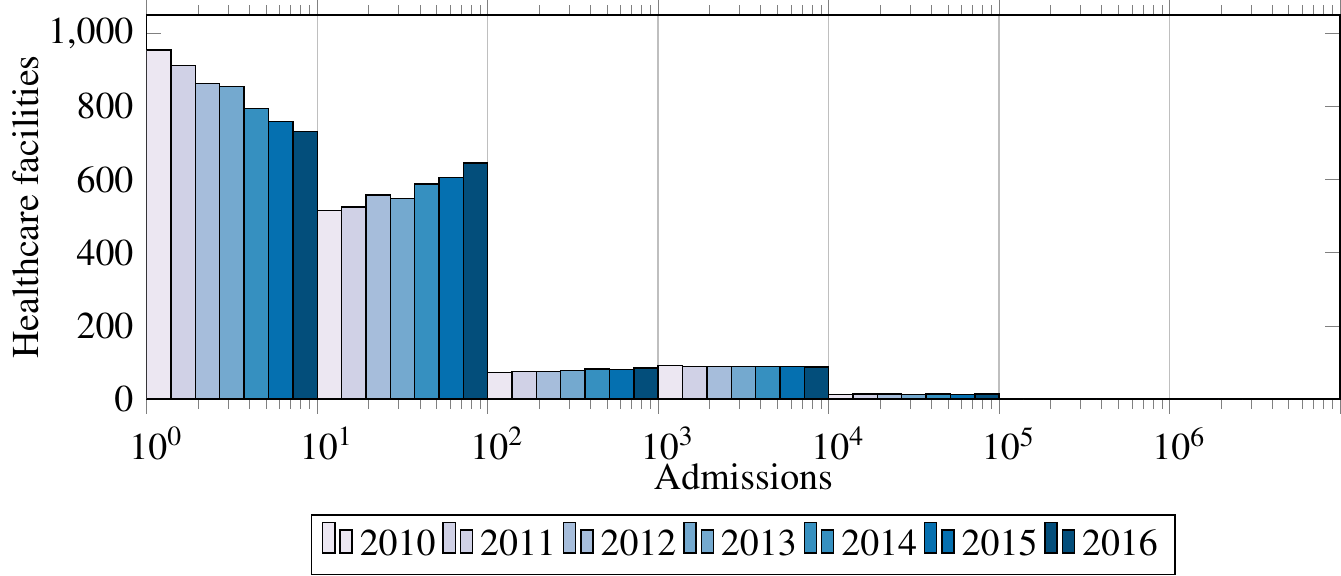}
%		\begin{tikzpicture}
%		\begin{axis}[
%		ybar interval,
%		width = \textwidth, height = 0.4\textwidth,
%		legend style={at={(0.5,-0.30)},anchor=north,legend columns=-1},
%		ymin=0,
%		xmin=1,
%		xmax=10e6,
%		xmode=log,
%		xlabel={Admissions},
%		ylabel={Healthcare facilities},
%		point meta=rawx,
%		xticklabel style={xshift=-2em},
%		]
%		\addplot[fill=blues1,] table [col sep=semicolon] {wyniki_text/hospitals_entriesno/y2010_hospitals_entriesno2.csv};
%		\addplot[fill=blues2] table [col sep=semicolon] {wyniki_text/hospitals_entriesno/y2011_hospitals_entriesno2.csv};
%		\addplot[fill=blues3] table [col sep=semicolon] {wyniki_text/hospitals_entriesno/y2012_hospitals_entriesno2.csv};
%		\addplot[fill=blues4]  table [col sep=semicolon] {wyniki_text/hospitals_entriesno/y2013_hospitals_entriesno2.csv};
%		\addplot[fill=blues5] table [col sep=semicolon] {wyniki_text/hospitals_entriesno/y2014_hospitals_entriesno2.csv};
%		\addplot[fill=blues6] table [col sep=semicolon] {wyniki_text/hospitals_entriesno/y2015_hospitals_entriesno2.csv};
%		\addplot[fill=blues7] table [col sep=semicolon] {wyniki_text/hospitals_entriesno/y2016_hospitals_entriesno2.csv};
%		\legend{2010, 2011, 2012, 2013, 2014, 2015, 2016};
%		\end{axis}
%		\end{tikzpicture}
		\caption{\label{fig:hosp:entries:all:b}}
	\end{subfigure}
	\caption{Number of healthcare facilities having given number of admissions for all cases reported: (a) within years 2010-2016, (b) for separate years.\label{fig:hosp:entries:all}}
\end{figure}

First, we characterize the healthcare facilities reported in the database by the number of admissions. Figure~\ref{fig:hosp:entries:all:a} shows that the majority healthcare facilities had between 10 and 99 admissions during years 2010--2016. This group is more than twice the size of the facilities with 10--999 reported admissions and almost three times the size of the facilities with 1--9 admissions only. The healthcare facilities with more than 1\,000 admissions were considerably less common. Clearly, units that have less than 10 patients during seven-year period might not be taken into account for the dynamical studies as their contribution to the patient transfer is insignificant. However, one should be aware that within particular years healthcare facilities with 1 to 9 admissions were the most numerous group (compare with Figure~\ref{fig:hosp:entries:all:b}).
This behaviour is due to the fact that these facilities are localized mostly in regions, which are covered by different insurance companies and thus data are very limited, often too broad to get accurate transfer dynamics.

\begin{figure}
	\centering
	\begin{subfigure}[t]{\textwidth}
				\includegraphics[width=0.9\linewidth]{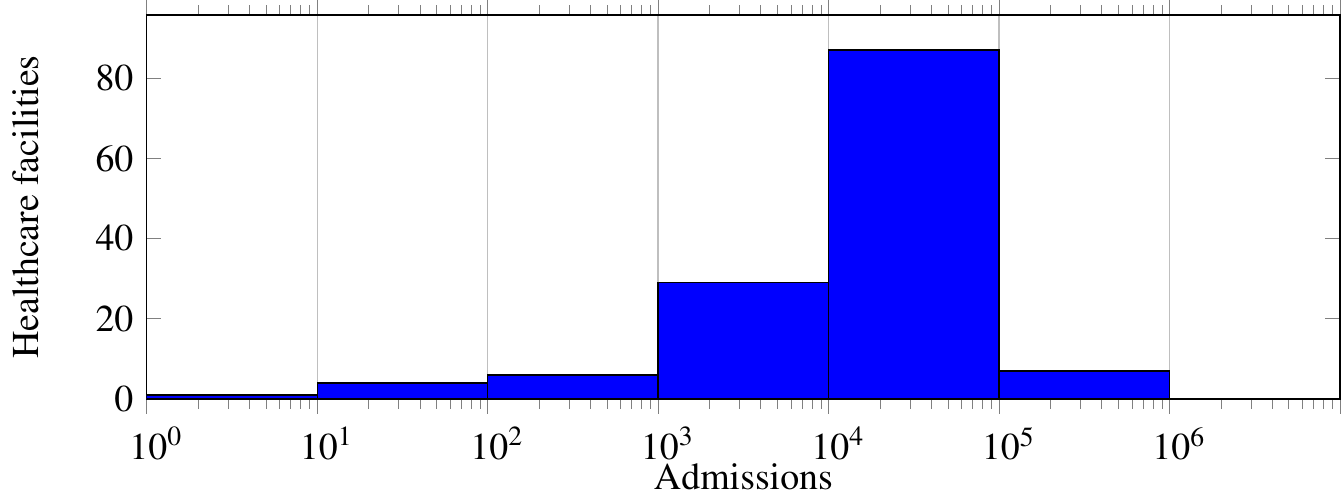}
%		\begin{tikzpicture}
%		\begin{axis}[
%		ybar interval,
%		width = \textwidth, height = 0.4\textwidth,
%		legend style={at={(0.5,-0.30)},anchor=north,legend columns=-1},
%		ymin=0,
%		xmin=1,
%		xmax=10e6,
%		xmode=log,
%		xlabel={Admissions},
%		ylabel={Healthcare facilities},
%		point meta=rawx,
%		xticklabel style={xshift=-2em},
%		]
%		\addplot[fill=blue] table [col sep=semicolon] {wyniki_text/hospitals_entriesno/st_hospitals_entriesno2.csv};
%		\end{axis}
%		\end{tikzpicture}
		\caption{\label{fig:hosp:entries:st:a}}
	\end{subfigure}
	~
	\begin{subfigure}[t]{\textwidth}
			\includegraphics[width=0.9\linewidth]{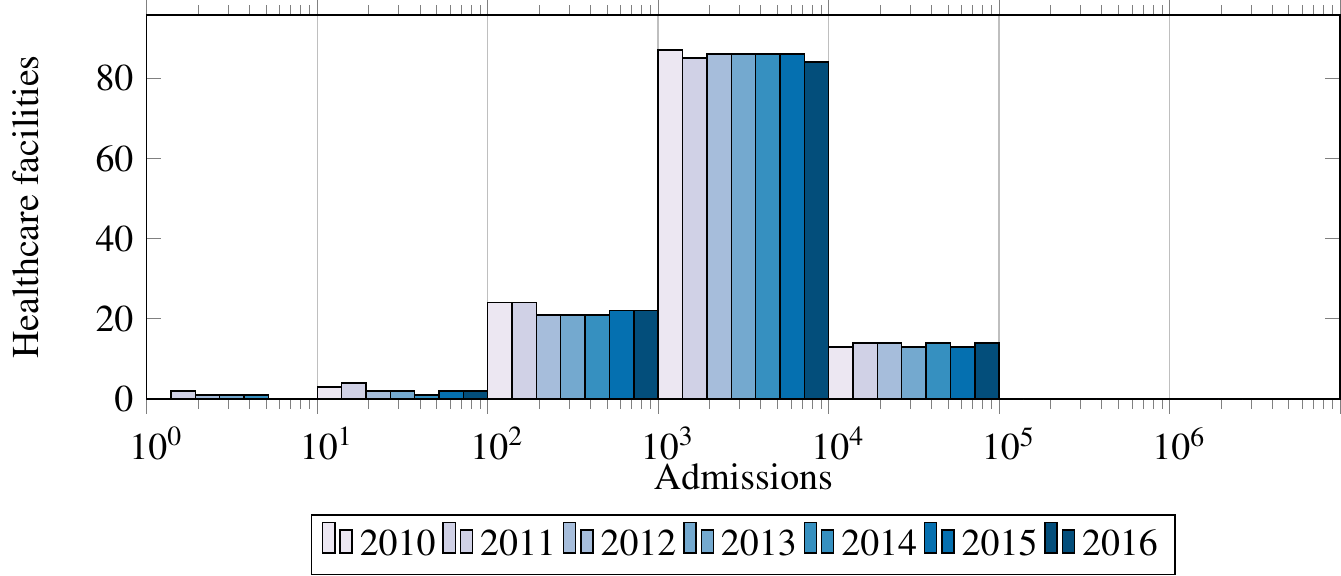}
%		\begin{tikzpicture}
%		\begin{axis}[
%		ybar interval,
%		width = \textwidth, height = 0.4\textwidth,
%		legend style={at={(0.5,-0.30)},anchor=north,legend columns=-1},
%		ymin=0,
%		xmin=1,
%		xmax=10e6,
%		xmode=log,
%		xlabel={Admissions},
%		ylabel={Healthcare facilities},
%		point meta=rawx,
%		xticklabel style={xshift=-2em},
%		]
%		\addplot[fill=blues1,] table [col sep=semicolon] {wyniki_text/hospitals_entriesno/st_y2010_hospitals_entriesno2.csv};
%		\addplot[fill=blues2] table [col sep=semicolon] {wyniki_text/hospitals_entriesno/st_y2011_hospitals_entriesno2.csv};
%		\addplot[fill=blues3] table [col sep=semicolon] {wyniki_text/hospitals_entriesno/st_y2012_hospitals_entriesno2.csv};
%		\addplot[fill=blues4]  table [col sep=semicolon] {wyniki_text/hospitals_entriesno/st_y2013_hospitals_entriesno2.csv};
%		\addplot[fill=blues5] table [col sep=semicolon] {wyniki_text/hospitals_entriesno/st_y2014_hospitals_entriesno2.csv};
%		\addplot[fill=blues6] table [col sep=semicolon] {wyniki_text/hospitals_entriesno/st_y2015_hospitals_entriesno2.csv};
%		\addplot[fill=blues7] table [col sep=semicolon] {wyniki_text/hospitals_entriesno/st_y2016_hospitals_entriesno2.csv};
%		\legend{2010, 2011, 2012, 2013, 2014, 2015, 2016};
%		\end{axis}
%		\end{tikzpicture}
		\caption{\label{fig:hosp:entries:st:b}}
	\end{subfigure}
	\caption{Number of healthcare facilities having given number of admissions for Saxony and Thuringia: (a) within years 2010-2016, (b) for separate years.\label{fig:hosp:entries:st}}
\end{figure}

The situation is quite different if we consider only the healthcare facilities situated in Saxony and Thuringia. From Figure~\ref{fig:hosp:entries:st:a} it is clear that the majority of facilities had between 10\,000 and 99\,999 admissions and the other groups were less common. This distribution is more suitable for studying the interhospital transmission of infections, because the facilities with very few patients in seven years do not have much impact on migrations of patients. For separate years, most Saxony and Thuringia healthcare facilities had between 1\,000 and 9\,999 reported admissions and the number of facilities with given intervals of reported admission did not change much with time (cf.~Figure~\ref{fig:hosp:entries:st:b}). Comparison of data reported in Figure~~\ref{fig:hosp:entries:all}~and~\ref{fig:hosp:entries:st} suggest that for the purpose of the realistic modelling, one should rather consider the healthcare facilities located in the considered federal states only, while other data can be used to estimate the exchange of patients between the regions.
	 
\subsection{Numbers of patients}
\begin{figure}
	\centering
	\begin{subfigure}[t]{\textwidth}
			\includegraphics[width=0.9\linewidth]{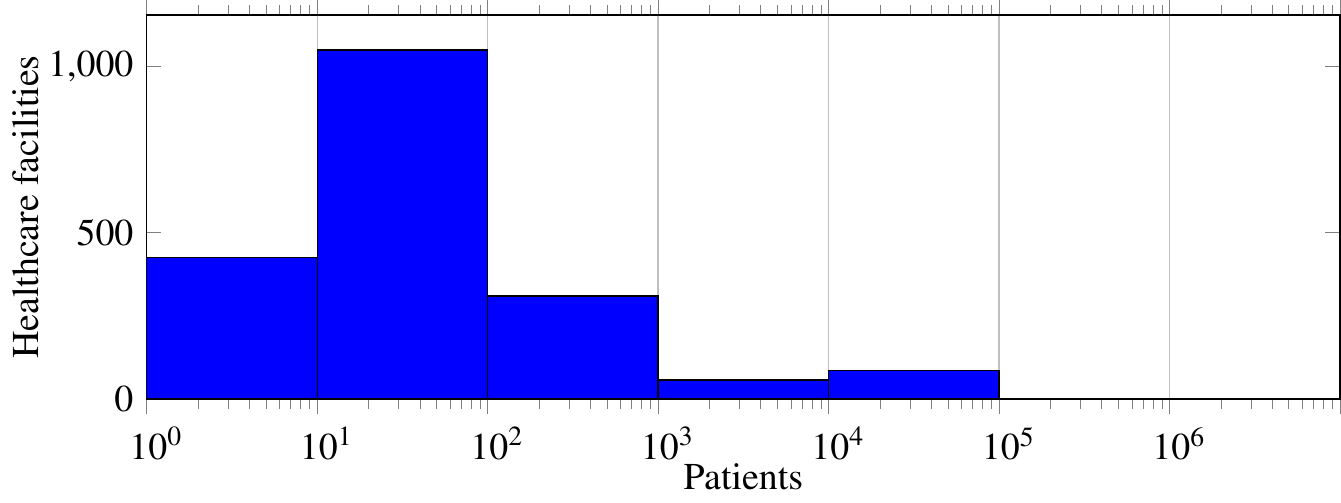}
%		\begin{tikzpicture}
%		\begin{axis}[
%		ybar interval,
%		width = \textwidth, height = 0.4\textwidth,
%		legend style={at={(0.5,-0.30)},anchor=north,legend columns=-1},
%		ymin=0,
%		xmin=1,
%		xmax=10e6,
%		xmode=log,
%		xlabel={Patients},
%		ylabel={Healthcare facilities},
%		point meta=rawx,
%		xticklabel style={xshift=-2em},
%		]
%		\addplot[fill=blue] table [col sep=semicolon] {wyniki_text/hospitals_patientsno/hospitals_patientsno2.csv};
%		\end{axis}
%		\end{tikzpicture}
		\caption{\label{fig:hosp:patients:all:a}}
	\end{subfigure}
	~
	\begin{subfigure}[t]{\textwidth}
			\includegraphics[width=0.9\linewidth]{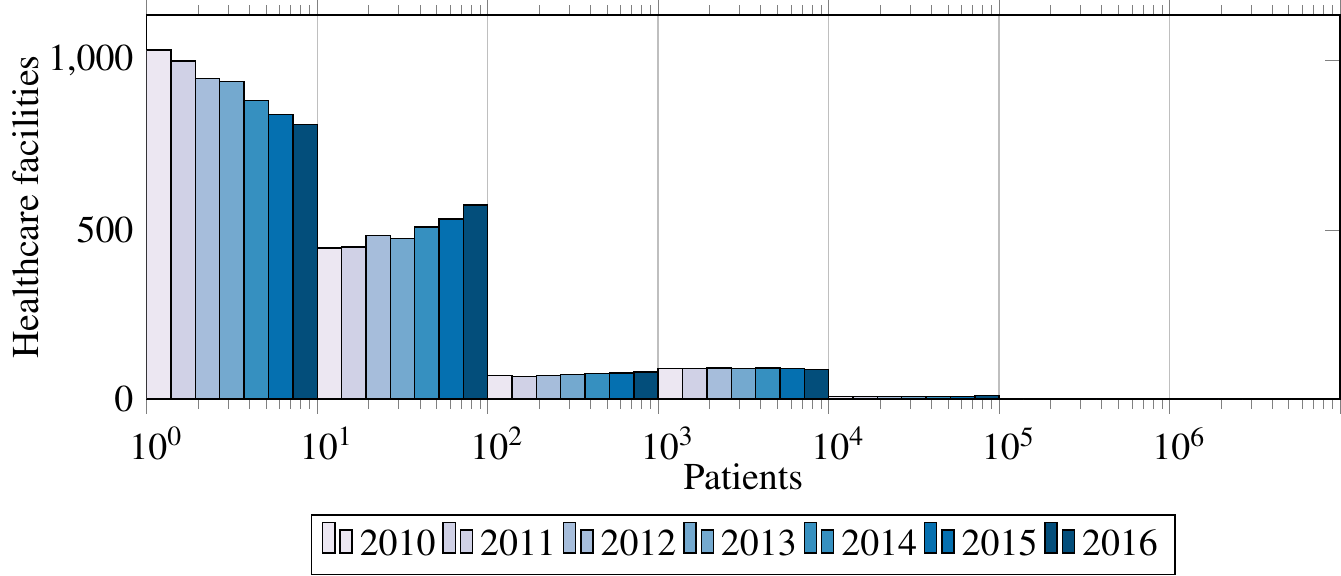}
%		\begin{tikzpicture}
%		\begin{axis}[
%		ybar interval,
%		width = \textwidth, height = 0.4\textwidth,
%		legend style={at={(0.5,-0.30)},anchor=north,legend columns=-1},
%		ymin=0,
%		xmin=1,
%		xmax=10e6,
%		xmode=log,
%		xlabel={Patients},
%		ylabel={Healthcare facilities},
%		point meta=rawx,
%		xticklabel style={xshift=-2em},
%		]
%		\addplot[fill=blues1,] table [col sep=semicolon] {wyniki_text/hospitals_patientsno/y2010_hospitals_patientsno2.csv};
%		\addplot[fill=blues2] table [col sep=semicolon] {wyniki_text/hospitals_patientsno/y2011_hospitals_patientsno2.csv};
%		\addplot[fill=blues3] table [col sep=semicolon] {wyniki_text/hospitals_patientsno/y2012_hospitals_patientsno2.csv};
%		\addplot[fill=blues4]  table [col sep=semicolon] {wyniki_text/hospitals_patientsno/y2013_hospitals_patientsno2.csv};
%		\addplot[fill=blues5] table [col sep=semicolon] {wyniki_text/hospitals_patientsno/y2014_hospitals_patientsno2.csv};
%		\addplot[fill=blues6] table [col sep=semicolon] {wyniki_text/hospitals_patientsno/y2015_hospitals_patientsno2.csv};
%		\addplot[fill=blues7] table [col sep=semicolon] {wyniki_text/hospitals_patientsno/y2016_hospitals_patientsno2.csv};
%		\legend{2010, 2011, 2012, 2013, 2014, 2015, 2016};
%		\end{axis}
%		\end{tikzpicture}
		\caption{\label{fig:hosp:patients:all:b}}
	\end{subfigure}
	\caption{Number of healthcare facilities having given number of patients for all cases reported: (a) within years 2010-2016, (b) for separate years.\label{fig:hosp:patients:all}}
\end{figure}

In order to estimate the probabilities of patient transfers between healthcare facilities,  size of the healthcare facilities is also necessary. Thus, we also categorise the healthcare facilities by the number of patients admitted to them. From Figure~\ref{fig:hosp:patients:all:a} we see that most healthcare facilities in all federal states had between 10 and 99 patients within years 2010--2016. Studying the data for each year separately, we conclude that for each year the facilities having between 1 and 10 patients dominated. However, the number of such facilities decreased over time and in 2016 the difference between their number and the number of the facilities having between 10 and 99 patients became less significant (cf.~Figure~\ref{fig:hosp:patients:all:b}).

\begin{figure}
	\centering
	\begin{subfigure}[t]{\textwidth}
			\includegraphics[width=0.9\linewidth]{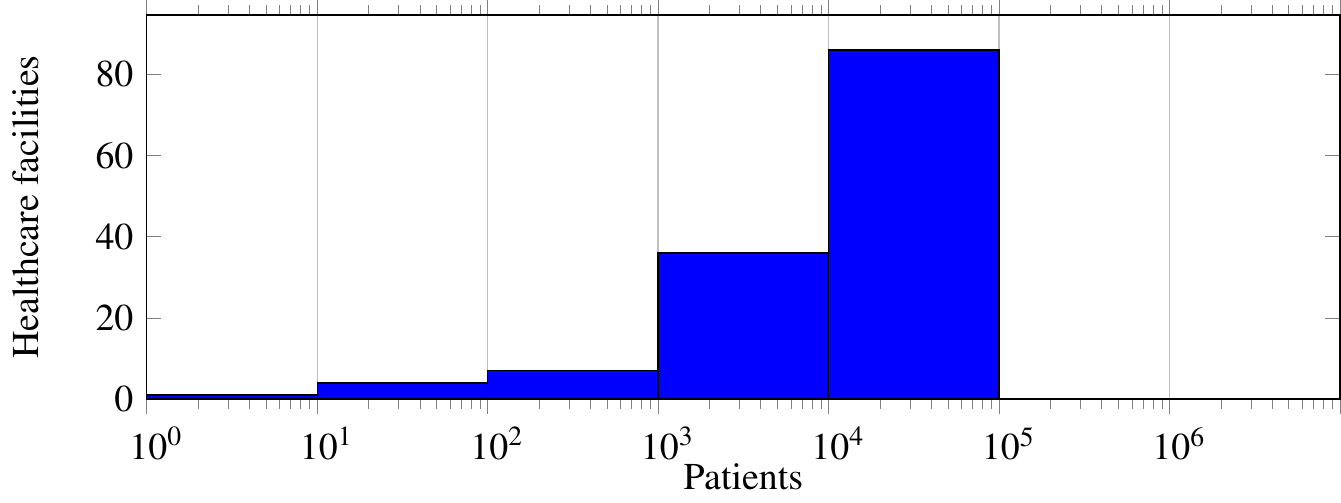}
%		\begin{tikzpicture}
%		\begin{axis}[
%		ybar interval,
%		width = \textwidth, height = 0.4\textwidth,
%		legend style={at={(0.5,-0.30)},anchor=north,legend columns=-1},
%		ymin=0,
%		xmin=1,
%		xmax=10e6,
%		xmode=log,
%		xlabel={Patients},
%		ylabel={Healthcare facilities},
%		point meta=rawx,
%		xticklabel style={xshift=-2em},
%		]
%		\addplot[fill=blue] table [col sep=semicolon] {wyniki_text/hospitals_patientsno/st_hospitals_patientsno2.csv};
%		\end{axis}
%		\end{tikzpicture}
		\caption{}
	\end{subfigure}
	~
	\begin{subfigure}[t]{\textwidth}
			\includegraphics[width=0.9\linewidth]{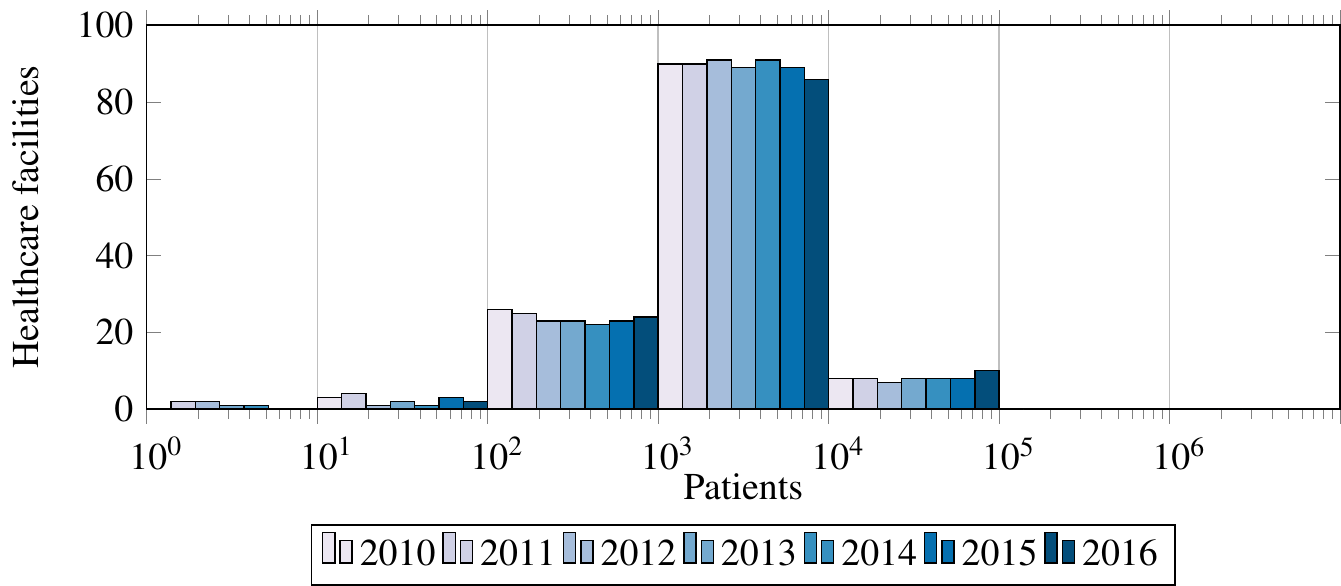}
%		\begin{tikzpicture}
%		\begin{axis}[
%		ybar interval,
%		width = \textwidth, height = 0.4\textwidth,
%		legend style={at={(0.5,-0.30)},anchor=north,legend columns=-1},
%		ymin=0,
%		xmin=1,
%		xmax=10e6,
%		xmode=log,
%		xlabel={Patients},
%		ylabel={Healthcare facilities},
%		point meta=rawx,
%		xticklabel style={xshift=-2em},
%		]
%		\addplot[fill=blues1,] table [col sep=semicolon] {wyniki_text/hospitals_patientsno/st_y2010_hospitals_patientsno2.csv};
%		\addplot[fill=blues2] table [col sep=semicolon] {wyniki_text/hospitals_patientsno/st_y2011_hospitals_patientsno2.csv};
%		\addplot[fill=blues3] table [col sep=semicolon] {wyniki_text/hospitals_patientsno/st_y2012_hospitals_patientsno2.csv};
%		\addplot[fill=blues4]  table [col sep=semicolon] {wyniki_text/hospitals_patientsno/st_y2013_hospitals_patientsno2.csv};
%		\addplot[fill=blues5] table [col sep=semicolon] {wyniki_text/hospitals_patientsno/st_y2014_hospitals_patientsno2.csv};
%		\addplot[fill=blues6] table [col sep=semicolon] {wyniki_text/hospitals_patientsno/st_y2015_hospitals_patientsno2.csv};
%		\addplot[fill=blues7] table [col sep=semicolon] {wyniki_text/hospitals_patientsno/st_y2016_hospitals_patientsno2.csv};
%		\legend{2010, 2011, 2012, 2013, 2014, 2015, 2016};
%		\end{axis}
%		\end{tikzpicture}
		\caption{}
	\end{subfigure}
	\caption{Number of healthcare facilities having given number of patients for Saxony and Thuringia: (a) within years 2010-2016, (b) for separate years.\label{fig:hosp:patients:st}}
\end{figure}

If we  consider the healthcare facilities located in Saxony and Thuringia only, during years 2010--2016 the facilities with 10\,000 to 99\,999 patients were the largest group. However, looking at the years separately, we see that for each year facilities having between 1\,000 and 9\,999 patients dominated by far (cf. Figure~\ref{fig:hosp:patients:st}).

\begin{figure}
	%\centerline{\includegraphics{.../......}} 
	\centering
		\begin{subfigure}[t]{0.5\textwidth}
				\includegraphics[width=0.9\linewidth]{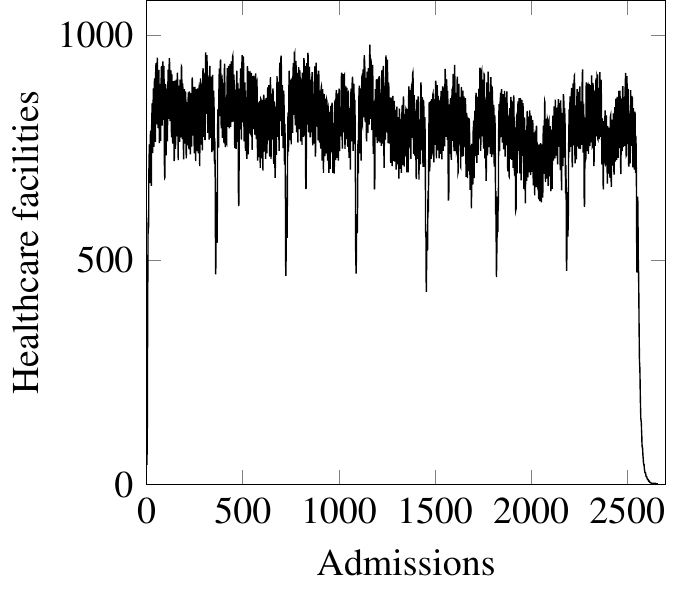}
%			\begin{tikzpicture}
%			\begin{axis}[
%			width = \textwidth, height = 6.5cm,
%			legend style={at={(0.5,-0.30)},anchor=north,legend columns=-1},
%			xmin=0,
%			ymin=0,
%			xmax=2700,
%			xlabel={Admissions},
%			ylabel={Healthcare facilities},
%			/pgf/number format/1000 sep={},
%			]
%			\addplot[mark=none] table [col sep = semicolon] {wyniki_text/days_patients_for_hosp/st_days_patients_for_hosp_02.csv};
%			\end{axis}
%			\end{tikzpicture}
			\caption{}
		\end{subfigure}%
	~ 
	\begin{subfigure}[t]{0.5\textwidth}
		\includegraphics[width=0.9\linewidth]{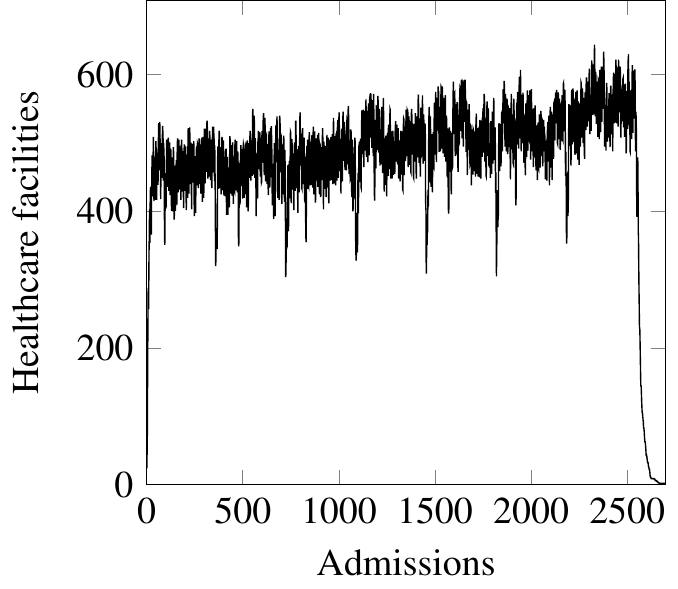}
%		\begin{tikzpicture}
%		\begin{axis}[
%		width = \textwidth, height = 6.5cm,
%		legend style={at={(0.5,-0.30)},anchor=north,legend columns=-1},
%		xmin=0,
%		ymin=0,
%		xmax=2700,
%		xlabel={Admissions},
%		ylabel={Healthcare facilities},
%		/pgf/number format/1000 sep={},
%		]
%		\addplot[mark=none] table [col sep = semicolon] {wyniki_text/days_patients_for_hosp/st_days_patients_for_hosp_12.csv};
%		\end{axis}
%		\end{tikzpicture}
		\caption{}
	\end{subfigure}  
	~
	\begin{subfigure}[t]{0.5\textwidth}
		\includegraphics[width=0.9\linewidth]{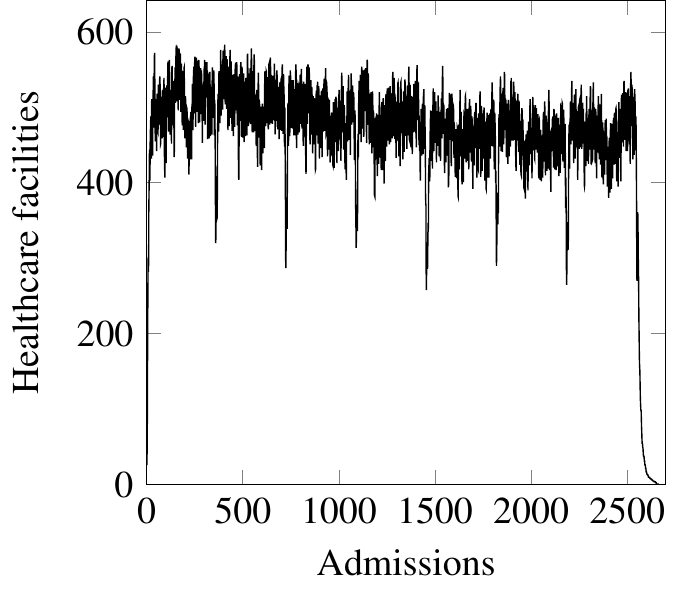}
%		\begin{tikzpicture}
%		\begin{axis}[
%		width = \textwidth, height = 6.5cm,
%		legend style={at={(0.5,-0.30)},anchor=north,legend columns=-1},
%		xmin=0,
%		ymin=0,
%		xmax=2700,
%		xlabel={Admissions},
%		ylabel={Healthcare facilities},
%		/pgf/number format/1000 sep={},
%		]
%		\addplot[mark=none] table [col sep = semicolon] {wyniki_text/days_patients_for_hosp/st_days_patients_for_hosp_22.csv};
%		\end{axis}
%		\end{tikzpicture}
		\caption{}
	\end{subfigure}%
	~ 
	\begin{subfigure}[t]{0.5\textwidth}
		\includegraphics[width=0.9\linewidth]{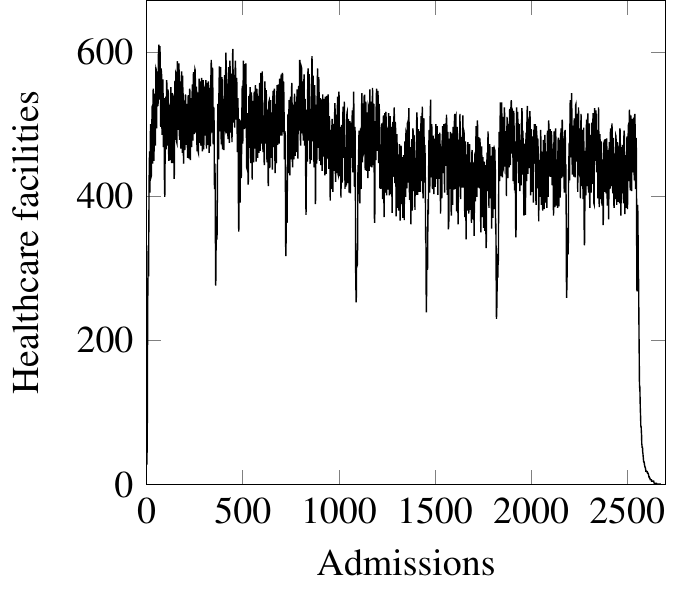}
%		\begin{tikzpicture}
%		\begin{axis}[
%		width = \textwidth, height = 6.5cm,
%		legend style={at={(0.5,-0.30)},anchor=north,legend columns=-1},
%		xmin=0,
%		ymin=0,
%		xmax=2700,
%		xlabel={Admissions},
%		ylabel={Healthcare facilities},
%		/pgf/number format/1000 sep={},
%		]
%		\addplot[mark=none] table [col sep = semicolon] {wyniki_text/days_patients_for_hosp/st_days_patients_for_hosp_32.csv};
%		\end{axis}
%		\end{tikzpicture}
		\caption{}
	\end{subfigure} 
	~
	\begin{subfigure}[t]{0.5\textwidth}
		\includegraphics[width=0.9\linewidth]{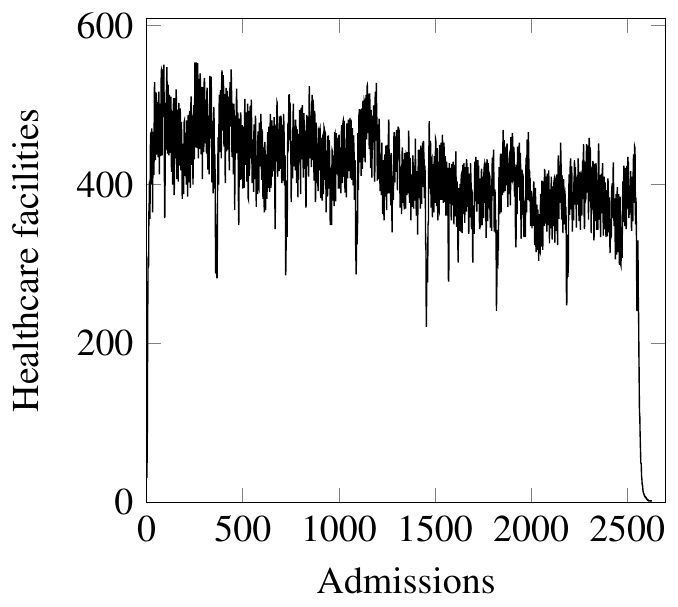}
%	\begin{tikzpicture}
%	\begin{axis}[
%	width = \textwidth, height = 6.5cm,
%	legend style={at={(0.5,-0.30)},anchor=north,legend columns=-1},
%	xmin=0,
%	ymin=0,
%	xmax=2700,
%	xlabel={Admissions},
%	ylabel={Healthcare facilities},
%	/pgf/number format/1000 sep={},
%	]
%	\addplot[mark=none] table [col sep = semicolon] {wyniki_text/days_patients_for_hosp/st_days_patients_for_hosp_42.csv};
%	\end{axis}
%	\end{tikzpicture}
		\caption{}
	\end{subfigure}%
	~ 
	\begin{subfigure}[t]{0.5\textwidth}
		\includegraphics[width=0.9\linewidth]{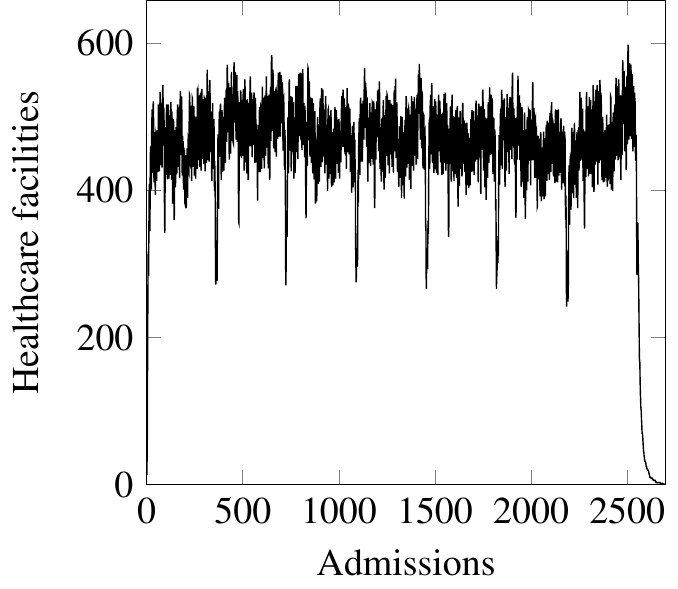}
%		\begin{tikzpicture}
%		\begin{axis}[
%		width = \textwidth, height = 6.5cm,
%		legend style={at={(0.5,-0.30)},anchor=north,legend columns=-1},
%		xmin=0,
%		ymin=0,
%		xmax=2700,
%		xlabel={Admissions},
%		ylabel={Healthcare facilities},
%		/pgf/number format/1000 sep={},
%		]
%		\addplot[mark=none] table [col sep = semicolon] {wyniki_text/days_patients_for_hosp/st_days_patients_for_hosp_52.csv};
%		\end{axis}
%		\end{tikzpicture}
		\caption{}
	\end{subfigure} 
	\caption{Number of patients staying in six biggest hospitals in Saxony and Thuringia within years 2010-2016.\label{fig:days:patients:all}}
\end{figure}

We analyse the changes of the number of patients of each healthcare facility in time (from 2010 up to 2016). In Figure~\ref{fig:days:patients:all} we present results for six biggest hospitals situated in Saxony and Thuringia.
In general, for the biggest hospitals, we can distinguish two kinds of processes. Clearly, there are periodic variations of the number of patients, which occur simultaneously among the healthcare system. On the other hand, there are long-term increase/decrease of the facility population's size, which are specific to healthcare facility.

\subsection{Duration of stays}
\begin{figure}
	\centering
	\begin{subfigure}[t]{0.5\textwidth}
			\includegraphics[width=0.9\linewidth]{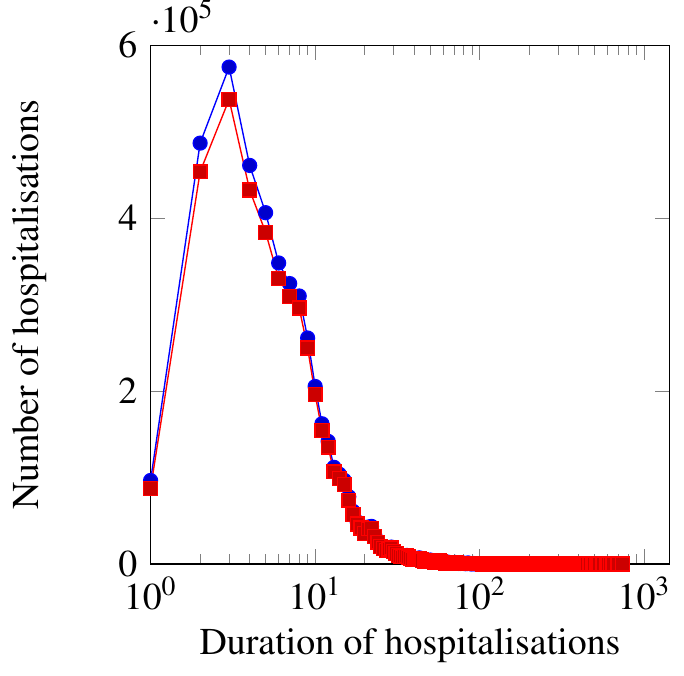}
%		\begin{tikzpicture}
%		\begin{axis}[
%		width = \textwidth, height = \textwidth,
%		legend style={at={(0.5,-0.30)},anchor=north,legend columns=-1},
%		ymin=0,
%		ymax = 600000,
%		xmin=1,
%		xmode =log,
%		ylabel={Number of hospitalisations},
%		xlabel={Duration of hospitalisations},
%		]
%		\addplot table [x index = 0, y index = 1, col sep=semicolon] {wyniki_text/entries_length/entries_length2.csv};
%		\end{axis}
%		\end{tikzpicture}
		\caption{\label{fig:durations:patients:a}}
	\end{subfigure}%
	~
	\begin{subfigure}[t]{0.5\textwidth}
		\includegraphics[width=0.9\linewidth]{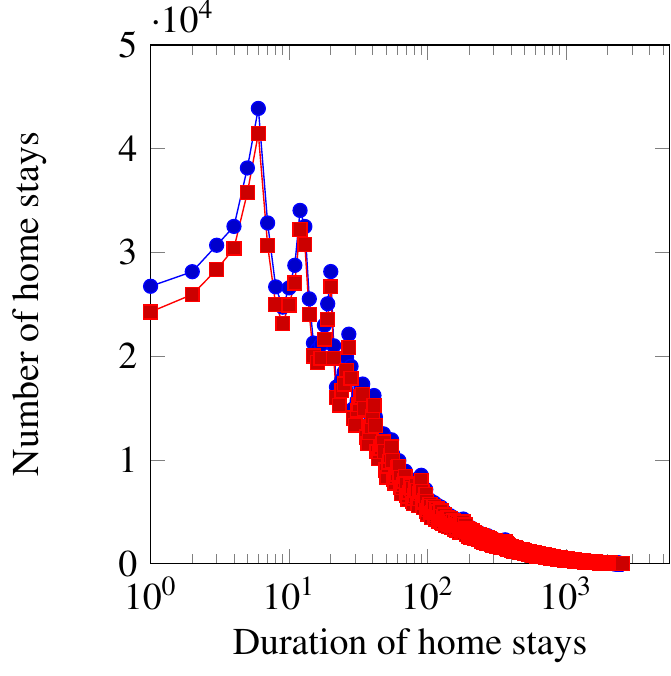}
%		\begin{tikzpicture}
%		\begin{axis}[
%		width = \textwidth, height = \textwidth,
%		legend style={at={(0.5,-0.30)},anchor=north,legend columns=-1},
%		ymin=0,
%		ymax = 600000,
%		xmin=1,
%		xmode =log,
%		ylabel={Number of hospitalisations},
%		xlabel={Duration of hospitalisations},
%		]
%		\addplot table [x index = 0, y index = 1, col sep=semicolon] {wyniki_text/entries_length/st_entries_length2.csv};
%		\end{axis}
%		\end{tikzpicture}
		\caption{\label{fig:durations:patients:b}}
	\end{subfigure}
	\caption{Durations of patients stays: (a) in all healthcare facilities (blue),  healthcare facilities located in Saxony and Thuringia (red) within years 2010-2016 (b) at home considering data for all healthcare facilities (blue) and healthcare facilities located in Saxony and Thuringia (red) within years 2010-2016.\label{fig:durations:patients}}
\end{figure}

We investigated duration of reported stays of patients in particular healthcare facilities. In Figure~\ref{fig:durations:patients:a} we present a histogram of the duration of the hospitalisations (until 31.12.2016) for all healthcare facilities (blue) and those for Saxony and Thuringia only (red) showing that there is no big difference between both datasets. We also observe the similar results comparing the lengths of stays at home between hospitalisations for all healthcare facilities in database and those located in Saxony and Thuringia, see Figure~\ref{fig:durations:patients:b}.

Clearly, majority of hospitalisations do not exceed 10 days and most of them are 3 days long. The number of hospitalisations quickly decreases for duration longer than 3 days. We see that hospitalisations that last at least a month constitute only a marginal part of all records. When it comes to stays at home between hospitalisations, the number of stays first increases as the duration growths, reaches maximum for six-day-long stays and then decreases with some fluctuations. However, the decline is considerably slower than for the hospitalisations and the stays lasting one hundred days or longer are still significant.

%\begin{figure}
%	\centering
%	\begin{subfigure}[t]{0.5\textwidth}
%%		\begin{tikzpicture}
%%		\begin{axis}[
%%		width = \textwidth, height = \textwidth,
%%		legend style={at={(0.5,-0.30)},anchor=north,legend columns=-1},
%%		ymin=0,
%%		ymax = 50000,
%%		xmin=1,
%%		xmode =log,
%%		ylabel={Number of home stays},
%%		xlabel={Duration of home stays},
%%		]
%%		\addplot table [x index = 0, y index = 1, col sep=semicolon] {wyniki_text/house_length/house_length2.csv};
%%		\end{axis}
%%		\end{tikzpicture}
%		\caption{}
%	\end{subfigure}%
%	~
%	\begin{subfigure}[t]{0.5\textwidth}
%%		\begin{tikzpicture}
%%		\begin{axis}[
%%		width = \textwidth, height = \textwidth,
%%		legend style={at={(0.5,-0.30)},anchor=north,legend columns=-1},
%%		ymin=0,
%%		ymax = 50000,
%%		xmin=1,
%%		xmode =log,
%%		ylabel={Number of home stays},
%%		xlabel={Duration of home stays},
%%		]
%%		\addplot table [x index = 0, y index = 1, col sep=semicolon] {wyniki_text/house_length/st_house_length2.csv};
%%		\end{axis}
%%		\end{tikzpicture}
%		\caption{}
%	\end{subfigure}
%	\caption{\agadd{Durations of patients stays at home considering data from: (a) all healthcare facilities, (b) healthcare facilities located in Saxony and Thuringia within years 2010-2016.\label{fig:home:patients}}}
%\end{figure}

\subsection{Overlaps}\label{sec:ovelaps}
Among 4\,826\,823 detected in the dataset hospitalisations we find 220\,367 cases of overlapping  records in total, while 
for healthcare facilities located in Saxony and Thuringia we find 198\,723 such cases.
By \emph{overlapping records} we understand distinct sets of two or more records for a given patient, with non-empty intersection of stay periods, either within the same facility or in other facilities.  
   
Following \cite{Piotrowska2019}, we distinguish several types of overlaps:
\begin{itemize}
	\item standard transfer --- one day overlap of two stay periods, where both periods are longer than one day and each record corresponds to different facility,
	\item first day transfer/last day transfer --- similar to above, but duration of the stay in one facility is exactly one day long and it coincides with admission to/discharge from the latter facility,
	\item simultaneous two admissions in a single institution --- two reported stays in the same place for the same period,
	\item temporary transfer  --- two records, period of one of them is contained in the other, and admission and discharge dates are not the same,
	\item simultaneous two admissions in two different institutions  --- periods are exactly the same, but the facilities are different,
	\item unknown two admissions in two different institutions --- any two records for hospitalisations in different institutions, which is not covered by the cases already introduced,
	\item two admissions in a single institution --- two reported stays in the same institution but for different (overlapping) periods,
	\item unknown multiple admissions ($n$) --- more than two records of overlapping hospitalisation periods, with maximal number of records in a given day is $n$.
\end{itemize}
In Figure~\ref{fig:examp} we present an exemplary visualization of the overlapping appearing in the database according to the proposed definitions.

\begin{figure}
{\footnotesize 
	%\scriptsize
	\begin{verbatim}
	
	47: | 2012-08-13: #########        |
	79: | 2012-08-13:         ######## |
	
	12: | 2010-05-24: ######## |
	24: | 2010-05-24: #        |

	5:  | 2016-07-08: #### |
	28: | 2016-07-08:    # |
	
	171: | 2013-09-01: #################### |
	193: | 2013-09-01:     ########         |
	
	19: | 2014-09-19: #################################        |
	72: | 2014-09-19:                      ################### |
	
	3: | 2014-01-28: ################### |
	3: | 2014-01-28:          ##         |
	8: | 2014-01-28:   ########          |
	
	49:  | 2010-09-04: ############        |
	182: | 2010-09-04: ################### |
	
	3:  | 2012-11-03: ##################### |
	12: | 2012-11-03: ##################### |	
		
	162: | 2011-03-11: ########### |
	162: | 2011-03-11: ########### |	
	
	7: | 2016-06-04: ##         |
	7: | 2016-06-04:  ######### |
	\end{verbatim}
}\caption{Examples of overlaps. In the first column the healthcare facility number is given, next initial date of hospitalisation and finally the graphical representation of hospitalisation duration (sign $\#$ denotes one day of stay in the healthcare facility). In the first row we see the example of {\it standard transfer}, next {\it first day transfer}, {\it last day transfer}, {\it temporary transfer}, {\it unknown two admissions in two institutions}, {\it unknown multiple admissions (3)},  another example of {\it unknown two admissions in two institutions}, {\it simultaneous two admissions in two institutions}, {\it simultaneous two admissions in a single institution, two admissions in a single institution}. \label{fig:examp}}
\end{figure} 

\begin{figure}
	\centering
	\begin{subfigure}[t]{\textwidth}
		\includegraphics[width=0.9\linewidth]{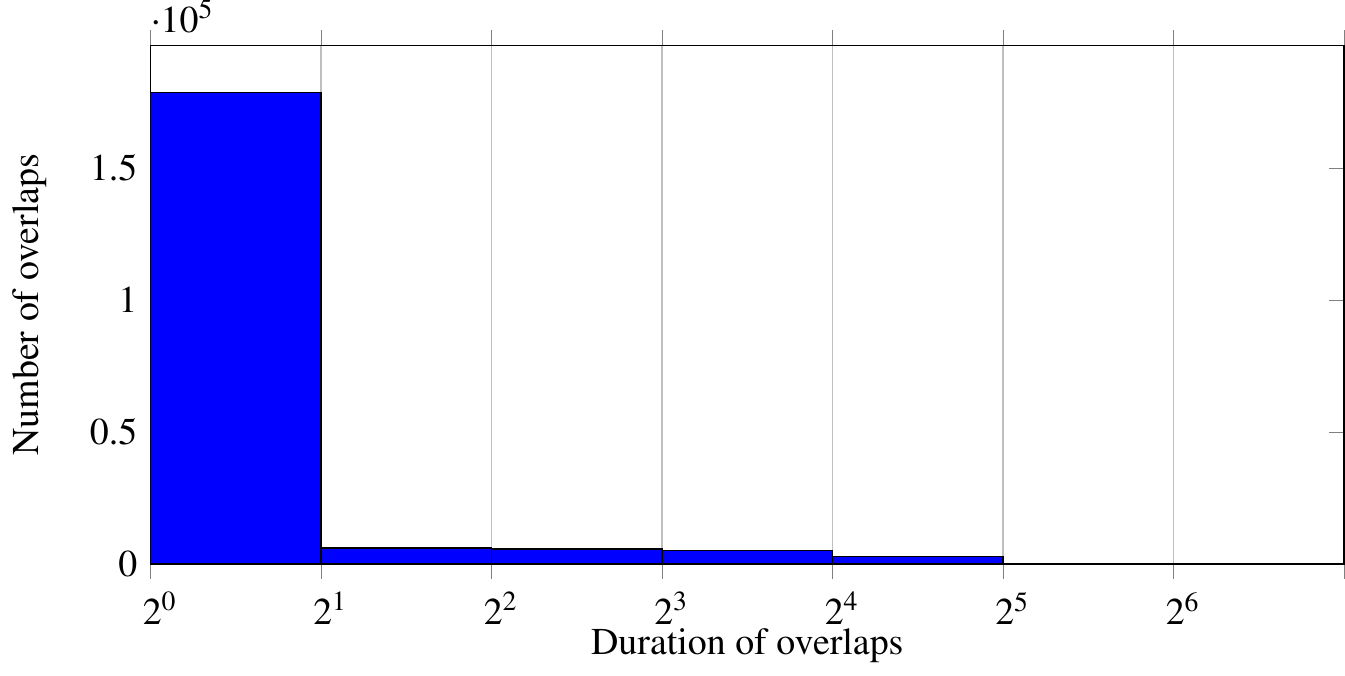}
%		\begin{tikzpicture}
%		\begin{axis}[
%		ybar interval,
%		width = \textwidth, height = 0.5\textwidth,
%		legend style={at={(0.5,-0.30)},anchor=north,legend columns=-1},
%		point meta=rawx,
%		ymin=0,
%		xmin=1,
%		xmax=128,
%		xmode = log,
%		log basis x={2},
%		xlabel={Duration of overlaps},
%		ylabel={Number of overlaps},		
%		xticklabel style={xshift=-2em},
%		]
%		\addplot[fill=blue] table [col sep=semicolon] {wyniki_text/overlaps_length/st_overlaps_length2.csv};
%		\end{axis}
%		\end{tikzpicture}
		\caption{}
	\end{subfigure}
	~
	\begin{subfigure}[t]{\textwidth}
				\includegraphics[width=0.9\linewidth]{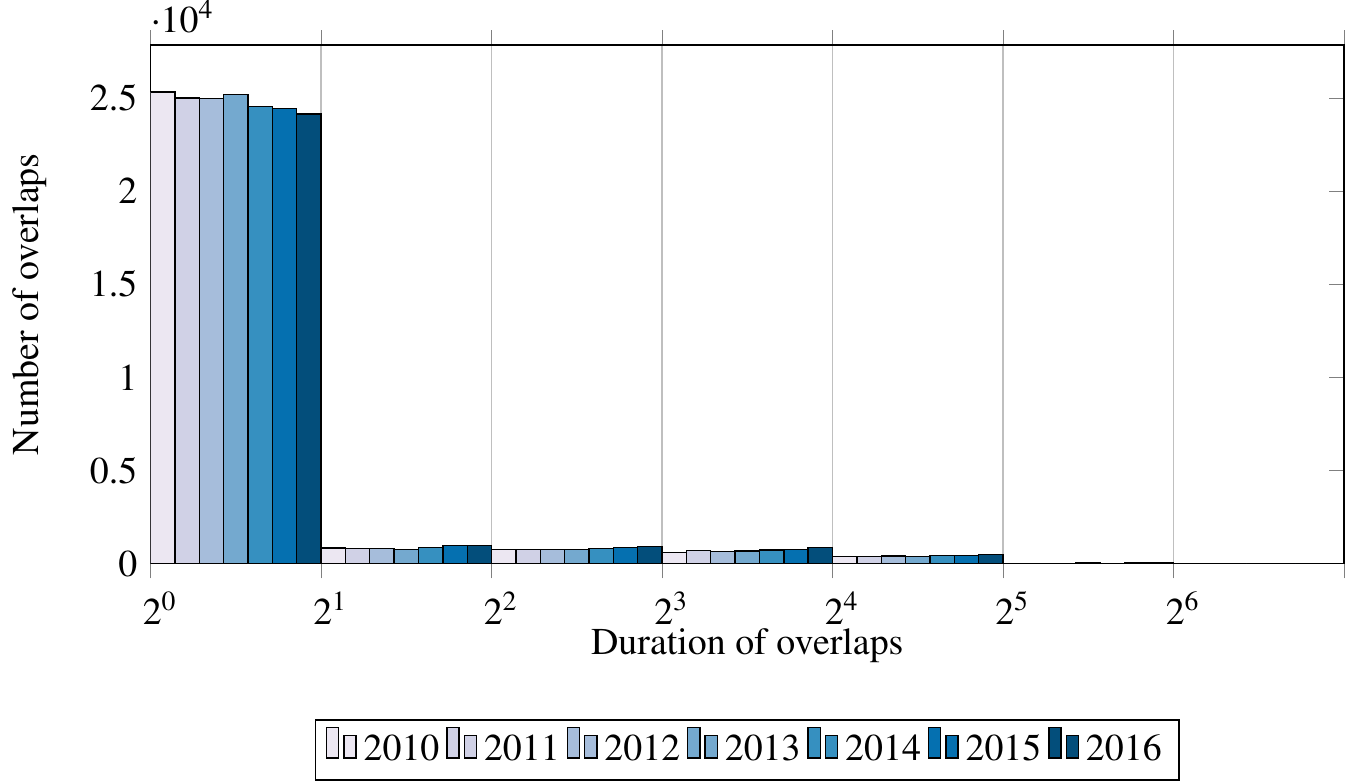}
%		\begin{tikzpicture}
%		\begin{axis}[
%		ybar interval,
%		width = \textwidth, height = 0.5\textwidth,
%		legend style={at={(0.5,-0.30)},anchor=north,legend columns=-1},
%		point meta=rawx,
%		ymin=0,
%		xmin=1,
%		xmax=128,
%		xmode = log,
%		log basis x={2},
%		xlabel={Duration of overlaps},
%		ylabel={Number of overlaps},	
%		xticklabel style={xshift=-2em},
%		]
%				\addplot[fill=blues1,] table [col sep=semicolon] {wyniki_text/overlaps_length/st_y2010_overlaps_length2.csv};
%				\addplot[fill=blues2] table [col sep=semicolon] {wyniki_text/overlaps_length/st_y2011_overlaps_length2.csv};
%				\addplot[fill=blues3] table [col sep=semicolon] {wyniki_text/overlaps_length/st_y2012_overlaps_length2.csv};
%				\addplot[fill=blues4]  table [col sep=semicolon] {wyniki_text/overlaps_length/st_y2013_overlaps_length2.csv};
%				\addplot[fill=blues5] table [col sep=semicolon] {wyniki_text/overlaps_length/st_y2014_overlaps_length2.csv};
%				\addplot[fill=blues6] table [col sep=semicolon] {wyniki_text/overlaps_length/st_y2015_overlaps_length2.csv};
%				\addplot[fill=blues7] table [col sep=semicolon] {wyniki_text/overlaps_length/st_y2016_overlaps_length2.csv};
%		\legend{2010, 2011, 2012, 2013, 2014, 2015, 2016};
%		\end{axis}
%		\end{tikzpicture}
		\caption{}
	\end{subfigure}
	\caption{Number of detected overlaps as a function of patient’s duration of stays for Saxony and Thuringia healthcare facilities reported: (a)~within years 2010-2016, (b)~in separate years.\label{fig:days:overlaps:st}}
\end{figure}

In Table~\ref{tab:overlaps} we present overlapping records for the whole database (all facilities, within years 2010--2016) and for units located in Saxony and Thuringia (the same time period) classified as described before. Comparing these two results we see that the structure of overlaps is almost the same for both sets. Namely, majority (over 77\%) of them are typical transfers, meaning that both stay periods are covered only by one day and the stays are reported for different institutions --- standard, first day and last day transfers. The other significant types are overlapping stays within one facility (over 12\%) and temporary transfers (over 7\%). Clearly, most problematic cases such as simultaneous admissions in three or more facilities are marginal and therefore can be ignored.
Generally the longer the overlaps are, the less often they appear in the database (cf. Figure~\ref{fig:days:overlaps:st}).

\begin{table}
	\centering
	%\aga{nowa tabela}
	\caption{Identified types of overlaps in AOK Plus database for the whole database and  for the units located in Saxony and Thuringia.}\label{tab:overlaps}
\begin{tabular}{|r|r|r|}
	\hline 
	\multirow{2}{*}{{\bf Overlap description}} & {\bf number of records}&{\bf number of records}\\
     &  {\bf whole database}  & {\bf Saxony and Thuringia}\\
	\hline
standard transfer & 152\,991 (69.4\%)& 137\,331 (69.1\%)\\ 	\hline 
two admissions in a single institution &  27\,625 (12.5\%)&  26\,402 (13.3\%)\\ 	\hline 
first day transfer &  16\,954 ( 7.7\%)&  14\,837 ( 7.5\%)\\ 	\hline 
temporary transfer &  16\,351 ( 7.4\%)&  14\,534 ( 7.3\%)\\ 	\hline 
unknown two admissions in two institutions &   3\,917 ( 1.8\%)&   3\,391 ( 1.7\%)\\ 	\hline 
unknown multiple admissions (3) &   1\,181 ( 0.5\%)&   1\,023 ( 0.5\%)\\ 	\hline 
last day transfer &    1\,000 ( 0.5\%)&    900 ( 0.5\%)\\ 	\hline 
simultaneous two admissions in two institutions &    308  ( 0.1\%)&    271  ( 0.1\%)\\ 	\hline 
simultaneous two admissions in a single institution &  34 (0.0\%)&  32 (0.0\%)\\ 	\hline 
unknown multiple admissions (4+) &     6 ( 0.0\%) &     2 ( 0.0\%)\\ 	\hline 
\end{tabular} 
\end{table}

%Our additional analysis shows that for Saxony and Thuringia 89.9\% (178\,618) 
%\mon{tu chyba ten sam bug, z transferami temp.-- akapit do poprawy} of the overlaps within years 2010--2016 last one day. Two- and three-day overlaps constitute respectively 1.6\% (3\,143) and 1.5\% (3\,012) of the overlaps.
%Generally the longer the overlaps are, the less often they appear in the database (cf. Figure~\ref{fig:days:overlaps:st}).
%Similar phenomenon occurs for the complete database, where one-day overlaps constitute 89.8\% (197\,864) of the overlaps within years 2010--2016.

In order to deeply characterize the types of overlaps, as in previous report~\cite{Piotrowska2019}, we use a four-digit classification. We restrict our analysis only to overlaps containing stays in one or two facilities. The truth is indicated by 1 while 0 means false. First digit indicate if two considered overlaps have place in the same healthcare facility, second digit: if overlaps have the same diagnoses, third: if two overlaps have the same admission dates and fourth if two overlaps have the same discharge dates. For example code 1100 simply means that two considered overlaps have been reported by the same healthcare facility, in both cases the diagnosis was the same, but there were different dates of admissions and discharges.

The results of our classification is presented in Table~\ref{tab:overlaps:codes} for all healthcare facilities and in Table~\ref{tab:overlaps:codes:st} for healthcare facilities located in Saxony and Thuringia. Clearly, the structure is the same for both datasets. The most frequently appearing overlap group is 0000 (different: facility, diagnose, admission and discharge date) -- 132\,170  detected cases for the whole dataset and 119\,045 cases for Saxony and Thuringia. Other quite frequent groups are 0100, 1000, 0010
(respectively: same diagnose, same facility, same admission dates) -- 38\,461, 23\,260, 13\,411 cases for the whole dataset and 34\,008, 22\,321, 11\,687 cases for Saxony and Thuringia, respectively.

Generally there are far more cases with different facilities than with the same one, which means there are more actual transfers. Cases where just one parameter out of the four considered is different are uncommon. It is also worth noting that there are no overlaps with code 1111, so there are no admissions, which look like repeated record.

\begin{table}
	\centering
	\caption{Effect of four-digit categorisation of the overlapping cases for all healthcare facilities.}\label{tab:overlaps:codes}
	\begin{tabular}{|l|c|c|c|c|c|c|c|c|}
		\hline
		{ {\bf overlap code}} & { 0000} & { 0100} & { 1000} & { 0010} & { 0110} & { 1100} & { 1010} & { 0001}\\
		\hline
		{ {\bf \# cases }} & { 132170} & { 38461} & { 23260} & { 13411} & { 6023} & { 2589} & { 1202} & { 967}\\
		\hline
		\hline
		{ {\bf overlap code}} & { 1001} & { 1110} & { 0011} & { 0101} & { 0111} & { 1101} & { 1011} & {1111} \\
		\hline
		{ {\bf \# cases }} & { 301} & { 209} & { 201} & { 181} & { 107} & { 64} & { 34} & {0}\\
		\hline
	\end{tabular}
\end{table}

\begin{table}
	\centering
	\caption{Effect of four-digit categorisation of the overlapping cases for Saxony and Thuringia healthcare facilities.}\label{tab:overlaps:codes:st}
	\begin{tabular}{|l|c|c|c|c|c|c|c|c|}
	\hline
	{ {\bf overlap code}} & { 0000} & { 0100} & { 1000} & { 0010} & { 0110} & { 1100} & { 1010} & { 0001}\\
	\hline
	{ {\bf \# cases }} & { 119045} & { 34008} & { 22321} & { 11687} & { 5225} & { 2406} & { 1133} & { 872}\\
	\hline
	\hline
	{ {\bf overlap code}} & { 1001} & { 1110} & { 0011} & { 0101} & { 0111} & { 1101} & { 1011} & {1111}\\
	\hline
	{ {\bf \# cases }} & { 288} & { 192} & { 176} & { 156} & { 95} & { 62} & { 32} & { 0}\\
	\hline
	\end{tabular}
\end{table}

Each of four-digit set we analyse further by assigning each diagnose into groups indexed
by numbers according to the rules presented in Table~3~in~\cite{Piotrowska2019} or \url{https://www.dimdi.de/static/de/klassifikationen/icd/icd-10-gm/kode-suche/htmlgm2019/}. In Tables~\ref{tab:overlaps:code:all}~and~\ref{tab:overlaps:code:st} we summarise the most frequently appearing diagnosis within the particular types of overlaps respectively for all healthcare facilities and for the ones in Saxony and Thuringia. From the presented results we see that most frequent diagnosis for two overlapping records related to different healthcare facilities is disease of the circulatory system [9, 9]. Other frequent problems are injuries [19, 19], neoplasms [2, 2] and mental disorders [5, 5].

For the overlaps characterised by the same healthcare facility records the vast majority of cases are [5, x], related to mental disorders. Among those the most frequent are cases [5, 19], which indicates change of diagnose from mental disorder to injury.

\begin{landscape}
\begin{table}
	\centering
\caption{Number of cases for a given diagnosis for all healthcare facilities for particular groups of overlaps. Two record overlaps are included in this table (vast majority among all overlaps). In square brackets we provide diagnose codes for overlaps versus number of cases.}\label{tab:overlaps:code:all}
\begin{tabular}{|l|r | |l|r | |l|r | |l|r | |l|r | |l|r | |l|r | |l|r | |}
	\hline
	{ 0000} & { Over.} & { 0001} & { Over.} & { 0010} & { Over.} & { 0011} & { Over.} & { 0100} & { Over.} & { 0101} & { Over.} & { 0110} & { Over.} & { 0111} & { Over.}\\
	\hline
	[9, 9]&24800 &[9, 9]&204 &[9, 9]&4667& [9, 9]&89&[9, 9]&15463&[9, 9]&91&[9, 9]&3603&[9, 9]& 78\\
	\hline
	[19, 19]&6932 &[5, 5]&84 &[19, 19]&1268 &[5, 5]&24&[19, 19]&7959&[5, 5]&24&[19, 19]&1085&[5, 5]&14\\
	\hline
	[2, 2]&6583 &[5, 19]&73 &[5, 5]&798&[5, 19]&11&[2, 2]&4510&[2, 2]&15&[15, 15]&203&[19, 19]&4\\
	\hline
	[5, 5]&5545&[9, 18]& 42 &[9, 19]&468&[19, 19]&9&[13, 13]&1725&[19, 19]&15&[11, 11]&171&[18, 18]&4\\
	\hline
	[5, 19]&4097 &[5, 9]&39&[14, 14]&454&[9, 18]&8&[5, 5]&1684&[11, 11] &9&[5, 5]&170&[11, 11]&2\\
	\hline
	[6, 9]&3506& [2, 2]&31 &[2, 2]&329&[6, 6]&5&[11, 11]&1353&[18, 18]&6&[14, 14]&150&[2, 2]&2\\
	\hline
	[9, 19]&3446 &[5, 18]&27 &[15, 15]&327&[6, 9]&5&[6, 6]&1144&[6, 6]&5&[2, 2]&120&[10, 10]&1\\
	\hline
	[13, 13]&3275 & [9, 11]&27&[6, 6]&323&[10, 10]&4&[10, 10]&1038&[13, 13]&4&[10, 10]&97& [14, 14]&1\\
	\hline
	[10, 10]&2966 &[6, 9]&26&[5, 19]&310&[2, 2]&4&[14, 14]&855&[10, 10]& 4&[18, 18]&95&[6, 6]&1\\
	\hline
	[5, 9]&2956& [9, 10]&24 &[6, 9]&304&[9, 19]&3&[1, 1]&508&[14, 14]&3&[6, 6]&86& { } & { }\\
	\hline
\end{tabular}

\begin{tabular}{|l|r | |l|r | |l|r | |l|r | |l|r | |l|r | |l|r ||  }
	\hline
	{ 1000} & { Over.} & { 1001} & { Over.} & { 1010} & { Over.} & { 1011} & { Over.} & { 1100} & { Over.} & { 1101} & { Over.} & { 1110} & { Over.} \\
	\hline
	[5, 19]&3353  & [5, 19]&62 &[5, 5]&328& [5, 19]&10&[5, 5]&1509&[5, 5]&46&[5, 5]&182\\
	\hline
	[5, 5]&2580 &[5, 5]&51&[5, 19]&241&[5, 5]&6&[6, 6]&386&[2, 2]&5&[15, 15]&8\\
	\hline
	[5, 9]&2448&[5, 9]&33 &[5, 6]&116&[5, 9]&5& [9, 9]& 236 & [9, 9]&3&[19, 19]&6 \\
	\hline
	[5, 6]&2311&[5, 18]& 28&[5, 18]&101&[5, 6]&3&[2, 2]&165& [11, 11]&2& [9, 9]&5 \\
	\hline
	[5, 11]&1460&[5, 11]&22&[5, 9]&75 &[16, 21]&1 &[15, 15] & 76&[19, 19]&2&[6, 6]&5 \\
	\hline
	[5, 18]&1372&[5, 6]&11&[4, 5]&64&[2, 5]&1&[19, 19]& 68&[6, 6]&2&[10, 10]&1 \\
	\hline
	[4, 5]&1057&[4, 5]&7&[5, 11]&60&[16, 16]&1& [11, 11]&34&[15, 15]&1&[14, 14]&1 \\
	\hline
	[5, 10]&883&[1, 5]&7&[5, 14]&25& [9, 19]&1 & [18, 18]&23&[14, 14]&1&[4, 4]&1\\
	\hline
	[2, 5]&702&[5, 10]&7&[1, 5]&21&[5, 10]&1&[10, 10]& 21& [16, 16]&1 & { } & { }  \\
	\hline
	[5, 14]&594&[5, 13]&6&[2, 5]&21&[5, 18]&1& [1, 1]&19 &[12, 12]&1& { } & { } \\
	\hline
\end{tabular}
\end{table}
\end{landscape}
\begin{landscape}
\begin{table}
	\centering
	\caption{Number of cases for a given diagnosis, for healthcare facilities in Saxony and Thuringia for particular groups of overlaps. Two record overlaps are included in this table (vast majority among all overlaps). In square brackets we provide diagnose codes for overlaps versus number of cases.}\label{tab:overlaps:code:st}
	\begin{tabular}{|l|r | |l|r | |l|r | |l|r | |l|r | |l|r | |l|r | |l|r | |}
		\hline
		{ 0000} & { Over.} & { 0001} & { Over.} & { 0010} & { Over.} & { 0011} & { Over.} & { 0100} & { Over.} & { 0101} & { Over.} & { 0110} & { Over.} & { 0111} & { Over.}\\
		\hline
	[9, 9]&22018&[9, 9]& 191&[9, 9]&4114&[9, 9]& 81& [9, 9]&13588 & [9, 9]&80 &[9, 9]&3124& [9, 9]&75\\
	\hline
	[19, 19]&6140 &[5, 19]& 69&[19, 19]&1070&[5, 5]&21& [19, 19]&7188 &[5, 5]&19 &[19, 19]&941&[5, 5]&11\\
	\hline
	[2, 2]&5955&[5, 5]& 68& [5, 5]&678 &[5, 19]&10& [2, 2]&4063 &[2, 2]&14&[15, 15]&184&[11, 11]&2\\
	\hline
	[5, 5]&4867&[9, 18]& 38&[9, 19]&424&[9, 18]& 8&[13, 13]&1547&[19, 19]&10 & [5, 5]&147 &[2, 2]&2\\
	\hline
	[5, 19]&3794 &[5, 9]&36 &[14, 14]&350&[19, 19]&6& [5, 5]&1404&[11, 11]&8&[11, 11]&146 &[19, 19]&2\\
	\hline
	[6, 9]&3255 &[2, 2]&30 & [2, 2]&306 &[6, 6]& 4 &[11, 11] &1199 &[18, 18]&6&[14, 14]&119&[18, 18]&2\\
	\hline
	[9, 19]&3103&[5, 18]&24&[6, 6]&297&[10, 10]&4&[6, 6]&1022 &[13, 13]&4&[2, 2]&105 &[10, 10]&1\\
	\hline
	[13, 13]&2979 &[6, 9]&24&[15, 15]&279&[6, 9]& 4&[10, 10]&924&[6, 6]&4& [10, 10]&89 & { } & { }\\
	\hline
	[5, 9]&2770 &[9, 11]&22&[6, 9]&273 &[2, 2]&3&[14, 14]&744 &[10, 10]&4&[18, 18]&85 & { } & { }\\
	\hline
	[10, 10]&2650&[9, 19]&19 &[5, 19]&266&[11, 11]&3&[1, 1]& 456 &[17, 17]&2& [6, 6]&80 & { } & { }\\
		\hline
	\end{tabular}
	
	\begin{tabular}{|l|r | |l|r | |l|r | |l|r | |l|r | |l|r | |l|r ||  }
		\hline
		{ 1000} & { Over.} & { 1001} & { Over.} & { 1010} & { Over.} & { 1011} & { Over.} & { 1100} & { Over.} & { 1101} & { Over.} & { 1110} & { Over.} \\
		\hline
		[5, 19]&3236 &[5, 19]&62&[5, 5]&305&[5, 19]&9 &[5, 5]&1384&[5, 5]&44 &[5, 5]&166\\
		\hline
		[5, 9]&2399 &[5, 5]&46 &[5, 19]&227&[5, 5]&6&[6, 6]&380&[2, 2]&5&[15, 15]&7\\
		\hline
		[5, 5]&2398 &[5, 9]&33 &[5, 6]&110&[5, 9]&5&[9, 9]&231&[9, 9]&3&[19, 19]&6\\
		\hline
		[5, 6]&2247 &[5, 18]&28&[5, 18]&93&[5, 6]&3 &[2, 2]&130&[11, 11]&2&[6, 6]&5 \\
		\hline
		[5, 11]&1400 &[5, 11]&21 &[5, 9]&74&[16, 21]&1&[15, 15]&72&[19, 19]&2&[9, 9]& 5 \\
		\hline
		[5, 18]&1308 &[5, 6]&11&[5, 11]&58&[2, 5]&1&[19, 19]&67&[6, 6]&2&[10, 10]&1 \\
		\hline
		[4, 5]&1008 &[4, 5]&7&[4, 5]&58&[9, 19]&1&[11, 11]&34&[15, 15]&1&[14, 14]&1 \\
		\hline
		[5, 10]&835 &[1, 5]&7&[5, 14]&25&[5, 10]&1 &[18, 18]&23&[14, 14]&1&[4, 4]&1\\
		\hline
		[2, 5]&672 &[5, 10]&7&[1, 5]&21&[5, 18]&1 &[10, 10]&19&[16, 16]&1& { } & { }  \\
		\hline
		[5, 14]&578 &[6, 9]&5&[2, 5]&21&[18, 18]&1&[1, 1]&18&[12, 12]&1& { } & { } \\
		\hline
	\end{tabular}
\end{table}
\end{landscape}

\section{Summary}\label{sec:summary}

In this paper we analysed the hospitalisation data provided by AOK Plus healthcare insurance company. First, the structure of population described in the dataset was examined. The population was stratificated by sex. Statistics concerning the number of hospitalisations and their lengths were computed. We also investigated the birth year of the patients and the diseases or health problems they were diagnosed with. We repeated the computations for records corresponding to facilities located in Saxony and Thuringia only and concluded that the subpopulation has very similar structure to the original one.

Then, we characterised the healthcare facilities by the number of admissions and the number of patients. We considered the dataset as a whole and when limited to Saxony and Thuringia region. For both groups we presented results for separate years and all years combined. We also found out that for the whole dataset small facilities (i.e. facilities with low number of admissions and patients) are the most numerous group thus, the whole dataset does not concern enough data to create base on them realistic hospital network well. However, if we consider Saxony and Thuringia only, most facilities are medium-sized or large, and the small ones are uncommon. Thus, we recommend for the modelling purpose to limit dataset to units located in Saxony and Thuringia.

Duration of hospitalisations and duration of home stays between hospitalisations were also analysed.
There are no significant differences in their distributions between the complete dataset and limited case.

The vital part of analysis regarding future modelling was investigating overlapping records. We introduced two types of classification. The first one describes, among others, number of healthcare facilities involved and length of the overlap. It shows that majority of the overlapping records are typical transfers between two different facilities for both whole dataset and subset of Saxony and Thuringia. Problematic cases, such as more than two facilities overlapping, are marginal. Additional analysis showed that overlaps lasting exactly one day overwhelmingly dominate over the longer ones. Another classification was introduced for overlaps involving no more than two hospitals. It describes whether the records have different facilities, diagnoses, admission and discharge dates. The structure is similar for both considered datasets. Overlaps involving two hospitals are far more frequent, which means there are more actual transfers between different facilities.
Overlaps containing one or two healthcare facilities were also briefly characterised by code representing the group of given diagnosis.

Analysis of the AOK Plus database is extremely useful for modelling the spread of e.g.~MDR bacterial infection. Information about population structure enables us to divide patients based on the risk factors, such as age or sex. Furthermore, the classification of healthcare facilities based on the number of admissions or the number of patients allows us to identify which facilities have sufficient data to be taken into account in the simulation. In particular, it is clear that such dataset alone is insufficient for simulation focused at different regions than Saxony and Thuringia, as the data is too scarce. Finally, the data concerning overlapping records can help us understand how certain factors impact on the patients' transitions.

\section{Acknowledgements}
This work was supported by grant no.~2016/22/Z/ST1/00690 of National Science Centre, Poland within the transnational research programme JPI-EC-AMR (Joint Programming Initiative on Antimicrobial Resistance) entitled "Effectiveness of infection control strategies against intra- and inter-hospital transmission of MultidruG-resistant Enterobacteriaceae –- insights from a multi-level mathematical NeTwork model" (EMerGe-Net). 

We thank the AOK Plus for providing anonymised record data. 
%\section{Appendix: Developed Python code}\label{app:code}
%
%{
%	\scriptsize
%	\begin{verbatim}
%		
%	\end{verbatim}
%	\lstinputlisting[language=Python]{program_po_utrechcie.py}

%%\section*{Bibliography}
\bibliographystyle{siam}
\bibliography{emergenet}

\begin{thebibliography}{1}

\bibitem{UErep2009}
{\em The bacterial challenge: time to react. {A} call to narrow the gap between
  multidrug-resistant bacteria in the {EU} and the development of new
  antibacterial agents.}, tech. rep., ECDC/EMEA,
  \url{https://ecdc.europa.eu/sites/portal/files/media/en/publications/Publications/0909_TER_The_Bacterial_Challenge_Time_to_React.pdf},
  2009.

\bibitem{InfectControlHospEpidemiol-2011-32-1064}
{\sc C.~Camus, E.~Bellissant, A.~Legras, A.~Renault, A.~Gacouin, S.~Lavou{\'e},
  B.~Branger, P.~Y. Donnio, P.~le~Corre, Y.~Le~Tulzo, D.~Perrotin, and
  R.~Thomas}, {\em {Randomized comparison of 2 protocols to prevent acquisition
  of methicillin-resistant Staphylococcus aureus: Results of a 2-Center study
  involving 500 patients}}, {Infection Control and Hospital Epidemiology}, 32
  (2011), pp.~1064--1072.

\bibitem{JAntimicrobChemother-2008-62-1422}
{\sc I.~F. Chaberny, F.~Schwab, S.~Ziesing, S.~Suerbaum, and P.~Gastmeier},
  {\em {Impact of routine surgical ward and intensive care unit admission
  surveillance cultures on hospital-wide nosocomial methicillin-resistant
  Staphylococcus aureus infections in a university hospital: An interrupted
  time-series analysis}}, {Journal of Antimicrobial Chemotherapy}, 62 (2008),
  pp.~1422--1429.

\bibitem{Donker2017}
{\sc T.~Donker, T.~Smieszek, K.~L. Henderson, A.~P. Johnson, A.~S. Walker, and
  J.~V. Robotham}, {\em Measuring distance through dense weighted networks: The
  case of hospital-associated pathogens}, {PLOS} Computational Biology, 13
  (2017), p.~e1005622.

\bibitem{Donker2012}
{\sc T.~Donker, J.~Wallinga, R.~Slack, and H.~Grundmann}, {\em Hospital
  networks and the dispersal of hospital-acquired pathogens by patient
  transfer}, {PLoS} {ONE}, 7 (2012), p.~e35002.

\bibitem{Lee2012}
{\sc B.~Y. Lee, S.~M. Bartsch, K.~F. Wong, S.~L. Yilmaz, T.~R. Avery, A.~Singh,
  Y.~Song, D.~S. Kim, S.~T. Brown, M.~A. Potter, R.~Platt, and S.~S. Huang},
  {\em Simulation shows hospitals that cooperate on infection control obtain
  better results than hospitals acting alone}, Health Affairs, 31 (2012),
  pp.~2295--2303.

\bibitem{AmerJPublHealth-2011-101-707}
{\sc B.~Y. Lee, S.~M. McGlone, Y.~Song, T.~R. Avery, S.~Eubank, C.~C. Chang,
  R.~R. Bailey, D.~K. Wagener, D.~S. Burke, R.~Platt, and S.~S. Huang}, {\em
  {Social network analysis of patient sharing among hospitals in Orange County,
  California.}}, {American Journal of Public Health}, 101 (2011), pp.~707--713.

\bibitem{NewEnglJMed-2010-362-1804}
{\sc A.~Y. Peleg and D.~C. Hooper}, {\em {Hospital-acquired infections due to
  gram-negative bacteria}}, {New England Journal of Medicine}, 362 (2010),
  pp.~1804--1813.

\bibitem{Piotrowska2019}
{\sc M.~J. Piotrowska and K.~Sakowski}, {\em Analysis of the {AOK} {L}ower
  {S}axony hospitalisation records data (years 2008 -- 2015)}, arXiv,
  1903.04701v1 (2019).

\end{thebibliography}

%\begin{thebibliography}{10}
%	\bibitem{Piotrowska2019}	
%  	{\sc M.~J. Piotrowska and K. Sakowski}, Analysis of the AOK Lower Saxony hospitalisation records data (years 2008--2015), \emph{arXiv:1903.04701}, (2019).
%
%%%	\bibitem{Banerjee2007}
%%%	{\sc S.~Banerjee and R.R. Sarkar}, {\em Delay-induced model for tumor-immune
%%%		interaction and control of malignant tumor growth}, BioSystems, 91 (2007),
%%%	pp.~268--288.
%%
%\end{thebibliography}

\end{document}